\documentclass[english,aps,prx,amsmath,amssymb,superscriptaddress,showpacs,superscriptaddress,floats,floatfix,nobalancelastpage,twocolumn]{revtex4-2}

\usepackage{graphicx}
\usepackage{dcolumn}
\usepackage{bm}
\usepackage{units}
\usepackage[dvipsnames]{xcolor}

\usepackage{hyperref}
\graphicspath{{./}{./figs/}{./psfigs/}{./Figures/}}

\newcommand{\cUS}[1]{\ensuremath{c_{#1}}}
\begin{document}

\title{Fractionalized Excitations Probed by Ultrasound}


\author{A. Hauspurg}
 \affiliation{Hochfeld-Magnetlabor Dresden (HLD-EMFL) and W\"urzburg-Dresden Cluster of Excellence ct.qmat, Helmholtz-Zentrum Dresden-Rossendorf (HZDR), 01328 Dresden, Germany}
\affiliation{Institut f\"ur Festk\"orper- und Materialphysik, Technische Universität Dresden, 01062 Dresden, Germany}

\author{S. Zherlitsyn}
\affiliation{Hochfeld-Magnetlabor Dresden (HLD-EMFL) and W\"urzburg-Dresden Cluster of Excellence ct.qmat, Helmholtz-Zentrum Dresden-Rossendorf (HZDR), 01328 Dresden, Germany}

\author{T. Helm}
\affiliation{Hochfeld-Magnetlabor Dresden (HLD-EMFL) and W\"urzburg-Dresden Cluster of Excellence ct.qmat, Helmholtz-Zentrum Dresden-Rossendorf (HZDR), 01328 Dresden, Germany}

\author{V. Felea}
\affiliation{Hochfeld-Magnetlabor Dresden (HLD-EMFL) and W\"urzburg-Dresden Cluster of Excellence ct.qmat, Helmholtz-Zentrum Dresden-Rossendorf (HZDR), 01328 Dresden, Germany}

\author{J. Wosnitza}
\affiliation{Hochfeld-Magnetlabor Dresden (HLD-EMFL) and W\"urzburg-Dresden Cluster of Excellence ct.qmat, Helmholtz-Zentrum Dresden-Rossendorf (HZDR), 01328 Dresden, Germany}
\affiliation{Institut f\"ur Festk\"orper- und Materialphysik, Technische Universität Dresden, 01062 Dresden, Germany}

\author{V. Tsurkan}
\affiliation{Experimental Physics V, Center for Electronic Correlations and Magnetism, University of Augsburg, 86135 Augsburg, Germany}
\affiliation{Institute of Applied Physics, MD 2028, Chisinau, Republic of Moldova}

\author{K.-Y. Choi}
\affiliation{Department of Physics, Sungkyunkwan University, Suwon 16419, Republic of Korea}
\author{S.-H. Do}
\affiliation{Materials Science and Technology Division, Oak Ridge National Laboratory, Oak Ridge, Tennessee 37831, USA}

\author{Mengxing Ye}
\affiliation{Kavli Institute for Theoretical Physics, University of California, Santa Barbara, CA 93106, USA}
\affiliation{Department of Physics and Astronomy, University of Utah, Salt Lake City, UT 84112, USA}

\author{Wolfram Brenig}
\affiliation{Institute for Theoretical Physics, Technical University Braunschweig, 38106 Braunschweig, Germany}

\author{Natalia B. Perkins}
\affiliation{School of Physics and Astronomy, University of Minnesota, Minneapolis, Minnesota 55455, USA}
\affiliation{Technical University of Munich, Germany, Institute for Advanced Study, D-85748 Garching, Germany}

\date{\today}

\begin{abstract}
In this work, we study magnetoelastic interactions by means of ultrasound experiments in $\alpha$-RuCl$_3$ -- a prototypical material for the Kitaev spin model on the honeycomb lattice, with a possible spin-liquid state featuring Majorana fermions and $\mathbb{Z}_{2}$-flux excitations. We present results of the temperature and in-plane magnetic-field dependence of the sound velocity and sound attenuation for several longitudinal and transverse phonon modes propagating along high-symmetry crystallographic directions. A comprehensive data analysis above the ordered state provides strong evidence of phonon scattering by Majorana fermions. This scattering depends sensitively on the value of the phonon velocities relative to the characteristic velocity of the low-energy fermionic excitations describing the spin dynamics of the underlying Kitaev magnet. 
Moreover, our data displays a distinct reduction of anisotropy of the sound attenuation, consistent with randomization, generated by thermally excited $\mathbb{Z}_2$ visons.
We demonstrate the potential of phonon dynamics
as a promising probe for uncovering fractionalized excitations in $\alpha$-RuCl$_3$ and provide new insights into the $H$-$T$ phase diagram of this material.
\end{abstract}

\maketitle

\section{Introduction}
A hallmark of a quantum spin liquid (QSL) is spin fractionalization.  In the exactly solvable Kitaev model spin excitations are fractionalized into two types of quasiparticles: itinerant spinon-like excitations, which are described by Majorana fermions, and localized gapped $\mathbb{Z}_2$ fluxes \cite{kitaev_anyons_2006, baskaran_exact_2007}.  Recent theoretical studies have shown that the characteristic signatures of such a spin fractionalization can be observed experimentally using various dynamical probes \cite{knolle_dynamics_2014,Knolle2015,Knolle2014raman,nasu2014vaporization,nasu2016fermionic,halasz2016resonant,halasz2019observing,Rousochatzakis2019,udagawa2021,Knolle2017,Motome2019, Trebst2022}.  Phonon dynamics is another indirect but promising probe to explore QSL physics in real materials \cite{Plee2011,Plee2013,metavitsiadis_phonon_2020,ye_phonon_2020,feng_temperature_2021,Metavitsiadis2022,Kexin2021}, since spin-lattice coupling is inevitable and often rather strong in real materials with large spin-orbit coupling.

The spin-orbit-coupled Mott insulator $\alpha$-RuCl$_3$ is among the most extensively studied materials predicted to host Kitaev interactions and to be in proximity to the Kitaev QSL state \cite{Plumb2014, banerjee_neutron_2017, do_majorana_2017, jansa_observation_2018,takagi_concept_2019,Trebst2022}. Despite the presence of non-Kitaev interactions in this material leading to a zig-zag antiferromagnetic order below $T_N$ = \unit[7.1]{K} \cite{Sears2015,balz_field_2021, suzuki_proximate_2021, wagner_magneto-optical_2022, bachus_thermodynamic_2020}, much effort has been devoted to searching for traces of fractionalization in the spin dynamics of $\alpha$-RuCl$_3$ \cite{li_identification_2021, laurell_dynamical_2020, modic_scale-invariant_2021, baek_evidence_2017,wolter_spin_2022, kasahara_majorana_2018, tanaka_thermodynamic_2022, yokoi_half-integer_2021, banerjee_neutron_2017}.  However, how close this material is to the Kitaev QSL remains an open question both at zero- and applied magnetic field \cite{bachus_angle_2021, bachus_thermodynamic_2020}.

It was realized that the magnetoelastic coupling plays an important role in the interpretation of thermal-Hall transport measurements ~\cite{kasahara_majorana_2018,ye2018quantization,vinkler-aviv_approximately_2018}. 
Therefore, the research focus has shifted further towards this aspect \cite{kocsis_magnetoelastic_2022, kaib_magnetoelastic_2021, schoenemann_thermal_2020,Mu2022} producing insights into the spin-phonon interactions in this material \cite{hentrich_unusual_2018, li_giant_2021, kasahara_majorana_2018, reschke_terahertz_2019, bruin_robustness_2022, yamashita_sample_2020}.
Mobile Majorana fermions and thermally induced static $\mathbb{Z}_2$ gauge fluxes lead to a continuum of scattering processes, which strongly affect the phonon dynamics and dissipation \cite{metavitsiadis_phonon_2020, ye_phonon_2020, feng_temperature_2021}.  In particular, low-temperature sound attenuation due to phonon scattering from fermionic particle-hole excitations is predicted to be linear in temperature \cite{metavitsiadis_phonon_2020, ye_phonon_2020}, and to show $C_{6v}$ symmetry \cite{ye_phonon_2020, feng_temperature_2021}.  Furthermore, time reversal symmetry breaking by a magnetic field results in a Hall viscosity, which mixes the longitudinal and transverse phonon modes, and allows for phonon Berry curvatures \cite{ye_phonon_2020}.

In this work, we report on a study of the spin dynamics in $\alpha$-RuCl$_3$ by means of ultrasound, examining the theoretical predictions related to the sound velocity and attenuation in the Kitaev material \cite{metavitsiadis_phonon_2020, ye_phonon_2020}. 
From the temperature and in-plane magnetic-field dependence of the attenuation of sound waves, we conclude that phonons are scattered off fractionalized excitations associated with the underlying Kitaev magnet. 
The difference in the sound attenuation of the longitudinal and transverse phonon modes 
can be understood by the
value of phonon  velocities with respect to the characteristic velocity of the low-energy fermionic excitations.
Remarkably, due to a significant difference in the velocities of the longitudinal and transverse phonon modes being larger and smaller than 
the  velocity of dispersing Majorana fermions, respectively, the phonon dynamics in $\alpha$-RuCl$_3$ exhibits two different  regimes of  attenuation: (1) a phonon scatters an occupied fermion state to a higher-energy state (ph-channel),
and (2) a
phonon decays  into two fermions, both
with positive energy (pp-channel).
Furthermore, thanks to the high sensitivity of ultrasound to various phase transitions 
\cite{luthi_physical_2005, truell_ultrasonic_1969}, we perform a detailed investigation of the $H$-$T$ phase diagram of $\alpha$-RuCl$_3$ for an in-plane magnetic field applied perpendicular to the Ru-Ru bonds.

\section{Methods}
We grow high-quality single crystals by vacuum sublimation \cite{reschke_sub-gap_2018, bachus_thermodynamic_2020}. Our samples show a single phase transition at \unit[7.1]{K} and no signatures of an additional phase transition at \unit[14]{K}, which could be caused by stacking faults.  While the trigonal $P3_112$ space group of $\alpha$-RuCl$_3$ has been reported at room temperature\ \cite{cao_low-temperature_2016}, the low-temperature symmetry of this compound is still under debate \cite{Mu2022,Lebert2022}.  In our investigation, we orient the samples using the Laue x-ray back-scattering diffraction technique. Analysis of the anisotropic suppression of magnetic order with in-plane magnetic fields confirms the sample orientation. There are two distinguished crystallographic directions in the honeycomb plane: the ${\bf a}$ axis perpendicular to the Ru-Ru bonds and the ${\bf b}$ axis parallel to the Ru-Ru bonds [see inset in Fig. \ref{fig:aRuCl3_c11_kIIa_Examples_Phasediagram_Tdep_Hdep}(c)].  The typical sample length in the honeycomb plane is \unit[2.0]{mm} along the {\bf a} and \unit[1.63]{mm} along the {\bf b} direction, with a characteristic size of $\unit[1]{mm}$ normal to the plane ({\bf c} direction).

We perform bulk ultrasound measurements using a pulse-echo method with a phase-sensitive detection technique \cite{zherlitsyn_2014, luthi_physical_2005}. To generate and detect ultrasonic signals in the frequency range of \unit[20 - 120]{MHz}, we bond ultrasound transducers (LiNbO$_3$, $36^{\circ}$ Y-cut for longitudinal and $41^{\circ}$ X-cut for transverse acoustic modes) with Thiokol 32 to the two parallel sample surfaces prepared by a focused ion beam (FIB).  Typically, several ultrasound echoes due to multiple propagations and reflections in the sample are observed in our experiments.  We use a calibrated RuO$_2$ resistor in a $^3$He cryostat and a Cernox sensor in a variable temperature insert (VTI) for thermometry.

In this work, we present results for the longitudinal and transverse acoustic modes, corresponding to the elastic stiffness constants \cUS{11}, and \cUS{33} as well as $(\cUS{11}-\cUS{12})/2$, and \cUS{44}, respectively, in a crystal assuming hexagonal symmetry (see Table \ref{tab:Schallgeschwindigkeiten} for the corresponding geometries). The related geometries, as well as the absolute values of the ultrasound frequency $f$, sound velocity $v_{s}$, and wave numbers $q$ are listed in Table \ref{tab:Schallgeschwindigkeiten}. For each of the acoustic modes, the general relationship, $\cUS{ij} = \rho v_{s}^2$, between the elastic constants, sound velocities, and mass density is satisfied. Note that as the symmetry of $\alpha$-RuCl$_3$ might be lower than hexagonal, i. e., $C2/m$  or $P3_112$, the sound velocities of acoustic modes along the ${\bf a}$ and ${\bf b}$ axes are different. In addition to the velocity, we characterize the dissipation of the acoustic-mode propagation by means of a sound-attenuation coefficient $\alpha$.  Because the initial signal amplitude $A(T_{0},H_{0})$ (measured at the initial field $H_0$ and temperature $T_0$) depends on the experimental settings, only the change of the sound attenuation is obtained from our experiments. We define this as $\Delta \alpha (T,H) = \alpha(T,H)-\alpha(T_0,H_0) = -20\log_{10}[A(T,H)/A(T_{0},H_{0})]/L$, where $A$ is the signal amplitude and $L$ is the effective sample length.

\begin{figure*}
	\centering
	\includegraphics{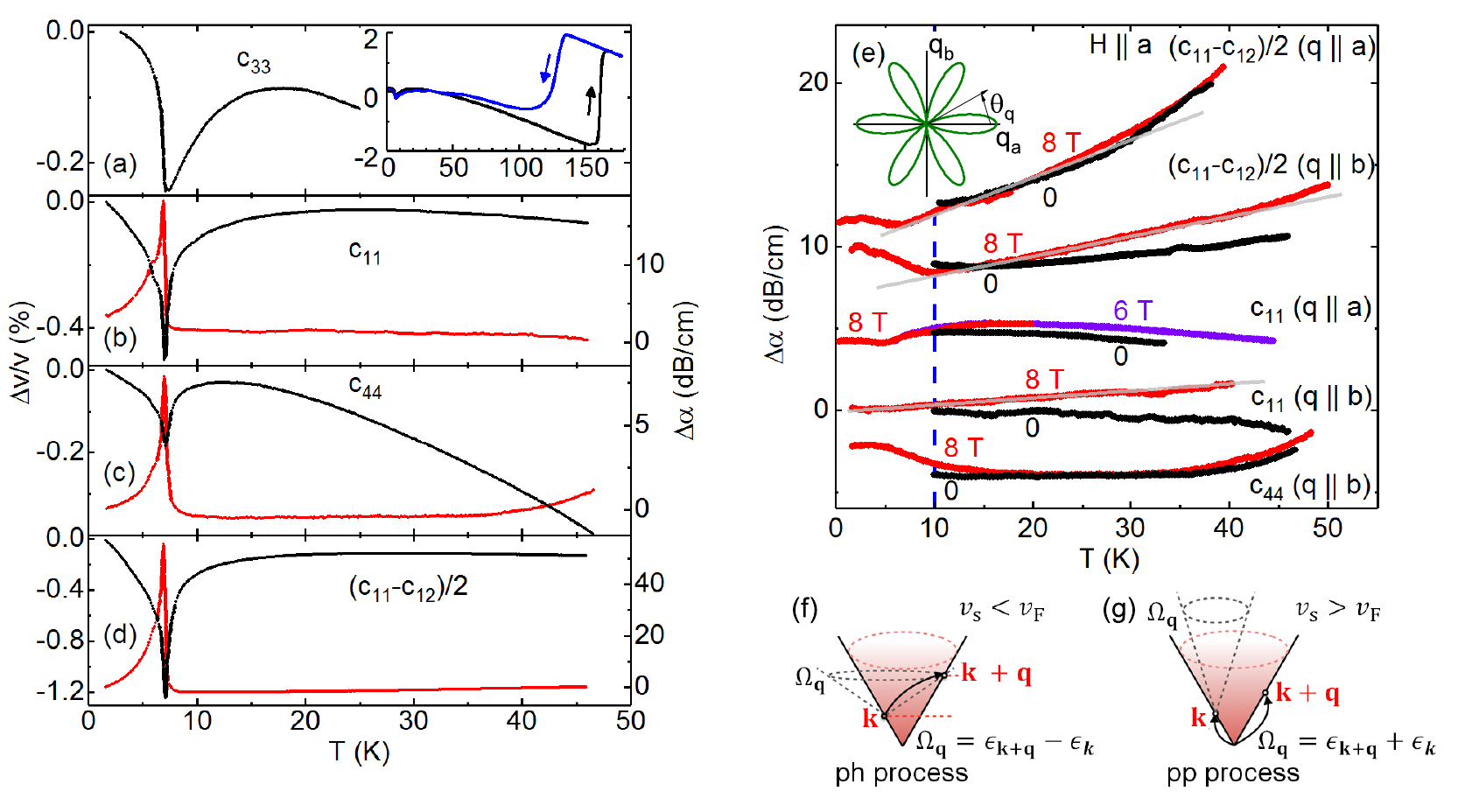}
	\caption{
	Temperature dependence of the relative sound velocity (black, left scales) and sound attenuation (red, right scales) of selected acoustic modes in $\alpha$-RuCl$_3$: (a) \cUS{33}, (b) \cUS{11}, (c) \cUS{44}, and (d) (\cUS{11}-\cUS{12})/2 with $\textbf{q} \parallel \textbf{b}$ in panels (b) to (d).
	Geometries and measurement frequencies are given in Table \ref{tab:Schallgeschwindigkeiten}.
	The inset in panel (a) shows the \cUS{33} up to \unit[180]{K} ($f=\unit[60]{MHz}$). Here, up and down sweeps are marked by arrows.
    (e) Change of the sound attenuation versus temperature in $\alpha$-RuCl$_3$ beyond the ordered state. We show results for the longitudinal \cUS{11} and transverse (\cUS{11}-\cUS{12})/2 acoustic modes for two high-symmetry in-plane propagation directions, $\textbf{q}\parallel \textbf{a}$ and $\textbf{q}\parallel \textbf{b}$, in zero (black) and magnetic field applied along {\bf a} (red, purple). Results for the transverse $\cUS{44}$ acoustic mode ($\textbf{q}\parallel \textbf{b}, \textbf{u}\parallel \textbf{c}$) are also shown. The zero-field data
    were removed close and below $T_N$ (below the vertical dashed blue line) to mask the dominant attenuation anomaly [shown in panels (a) to (d)] at the ordering temperature $T_N$. 
    The gray straight lines drawn to guide the eye highlight the linear temperature dependence of the sound attenuation above the ordered phase. Measurement frequencies were in the range of \unit[20 - 40]{MHz}. Note that for different experimental geometries (different acoustic modes and different propagation directions) the related attenuation is arbitrarily shifted for clarity. The zero and in-field curves belonging to the same acoustic mode for a chosen propagation direction [and with the same $A (T_{0},H_0)$] are not shifted with respect to each other. (f) and (g) illustrate  different scattering channels near the bottom of the Dirac cone with acoustic phonon velocities $v_{s} < v_{F}$ and $v_{s} > v_{F}$, respectively. See text for details.}
	\label{fig:Fig1_aRuCl3_0Field_Alpha_Cones}
\end{figure*}

\section{Results}
\subsection{Sound Velocity and Attenuation}
Next we present the main result of this paper - the experimental observation of a rich temperature and in-plane magnetic-field dependence of the sound attenuation beyond the ordered state in $\alpha$-RuCl$_3$.

Figure \ref{fig:Fig1_aRuCl3_0Field_Alpha_Cones}(a-d) shows the temperature dependence of the relative sound velocity, $\Delta v/ v$, and the change in the sound attenuation, $\Delta \alpha(T)$ in the different acoustic-mode geometries. To begin with, this figure clearly establishes the presence of a strong magnetoelastic coupling for sound propagation in $\alpha$-RuCl$_3$. The acoustic modes exhibit pronounced anomalies at the magnetic ordering temperature $T_N$, which are fully consistent with the critical softening due to a second-order magnetic transition \cite{Luthi1970}, displaying a sharp dip in $\Delta v/v (T)$ and a peak in $\Delta \alpha (T)$. The inset in Fig. \ref{fig:Fig1_aRuCl3_0Field_Alpha_Cones}(a) shows a thermal hysteresis of the $c_{33}$ mode at the first-order structural phase transition, presumably, from the high-temperature monoclinic to low-temperature rhombohedral structure \cite{Glamazda2017, Park2016}.
 
\begin{figure*}
	\centering
	\includegraphics{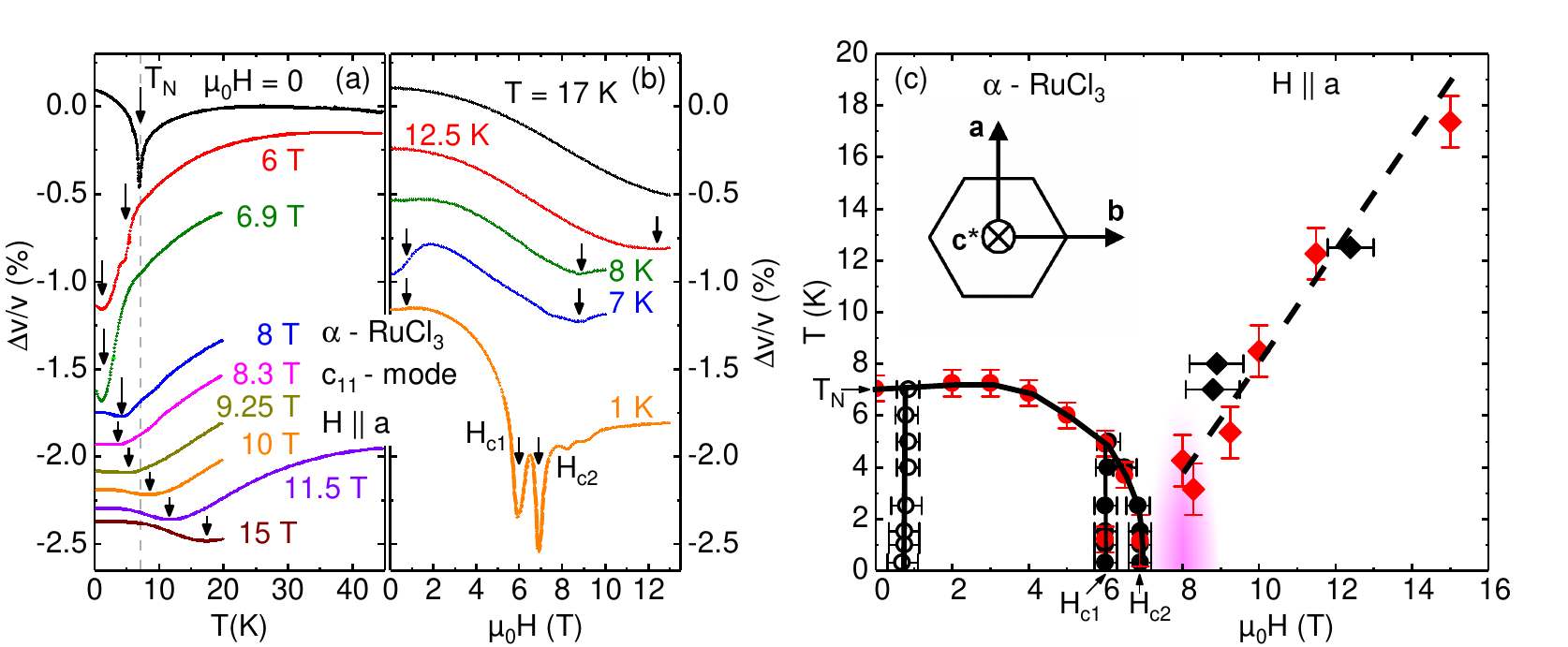}
	\caption{(a) Temperature and (b) magnetic field dependence of the relative sound velocity of the longitudinal acoustic mode \cUS{11} ($\textbf{q}\parallel \textbf{u}\parallel \textbf{a}$) at selected magnetic fields (${\bf H}\parallel \textbf{a}$) and temperatures, respectively. The curves are arbitrarily shifted for clarity. The measurement frequency was $f=\unit[22]{MHz}$. The vertical arrows indicate positions of the sound-velocity anomalies. $T_N$, $H_{c1}$, and $H_{c2}$ are also shown. The grey dashed line in (a) corresponds to $T_N$. (c) $H$-$T$ (${\bf H} \parallel \textbf{a}$) phase diagram of $\alpha$-RuCl$_3$. Characteristic temperatures (red symbols) and magnetic fields (black symbols) are extracted from the ultrasound data shown in (a) and (b). The lines are guides to the eye and the pink area marks the putative intermediate phase. According to \cite{kasahara_majorana_2018} this is the most likely regime for a QSL in $\alpha$-RuCl$_3$. The inset shows crystallographic directions in a honeycomb plane with the notations used in this work.}
	\label{fig:aRuCl3_c11_kIIa_Examples_Phasediagram_Tdep_Hdep}
\end{figure*}

In Fig. \ref{fig:Fig1_aRuCl3_0Field_Alpha_Cones}(e), we report the temperature dependence of the sound-attenuation change of the longitudinal, \cUS{11}, and transverse, $(\cUS{11}-\cUS{12})/2$ and \cUS{44}, acoustic modes measured along two high-symmetry in-plane propagation directions, $\textbf{q}\parallel \textbf{a}$ and $\textbf{q}\parallel \textbf{b}$, in zero and in finite magnetic field applied along the $\textbf{a}$ direction [see the inset of Fig. \ref{fig:aRuCl3_c11_kIIa_Examples_Phasediagram_Tdep_Hdep}(c)]. 
In the following, we will focus on three relevant aspects. First, for a given propagation direction and magnetic field, $\Delta\alpha(T)$ varies differently with $T$ for different modes. Second, for each given mode and magnetic field, $\Delta\alpha(T)$ depends on the in-plane propagation directions, i. e., $\textbf{a}$ versus $\textbf{b}$. Finally, each acoustic mode with in-plane strain exhibits different zero- and finite-field attenuation `slopes' versus temperature. We note that the temperature dependence of $\Delta\alpha(T)$ due to lattice anharmonicity should not be field dependent.

In the zero-field data, it is remarkable that $\Delta\alpha(T)$ not only shows a very different temperature dependence for the in-plane longitudinal, $\cUS{11}$, and the transverse, $(\cUS{11}-\cUS{12})/2$, acoustic modes, but it is also quite different for each of the two geometries, with respect to the same acoustic mode. Within the temperature range of \unit[10--35]{K}, we observe a clear linear temperature dependence of $\Delta\alpha(T)$ for the transverse modes $(\cUS{11}-\cUS{12})/2$ propagating along the $\textbf{q}\parallel \textbf{a}$ and $\textbf{q}\parallel \textbf{b}$ directions. Moreover, the slope for the transverse mode, $(\cUS{11}-\cUS{12})/2$, propagating along $\textbf{q}\parallel \textbf{a}$ is significantly larger than that propagating along $\textbf{q}\parallel \textbf{b}$. Conversely, the longitudinal mode, $\cUS{11}$, only shows an overall weak downward trend of $\Delta\alpha(T)$, which may slightly increase above \unit[20]{K}.

We rationalize these findings based on the fermionic excitations of the pure Kitaev model. We neglect thermal flux excitations initially and return to them later. The Fermi velocity, which is determined by the 
slope of the Dirac cones, can be estimated via the Kitaev coupling, $J\simeq \unit[81]{K}$ (\unit[7]{meV}), in $\alpha$-RuCl$_3$ to be $v_F \simeq \unit[2746]{m/s}$ (\unit[18]{meV$\cdot$\AA}). This is consistent with the previous estimates in  Ref. \cite{li_divergence_2021}. Apparently, from Table I, the sound velocities $v_{s}^{Ta}$ and $v_{s}^{Tb}$ of the transverse mode $(\cUS{11}-\cUS{12})/2$ for $\textbf{q}\parallel \textbf{a}$ and $\textbf{q}\parallel \textbf{b}$, respectively, are both smaller than $v_F$. In this situation, the primary contribution to the sound attenuation stems from the microscopic processes, in which a phonon scatters a positive-energy fermion to a 
higher-energy fermion state -- dubbed ph-channel, Fig. \ref{fig:Fig1_aRuCl3_0Field_Alpha_Cones}(f). These processes satisfy both energy and momentum-kinematic constraints. Power counting for the density of states in the vicinity of the Dirac points leads to $\Delta\alpha(T) \propto T$, i. e., linearity in temperature \cite{metavitsiadis_phonon_2020, ye_phonon_2020,feng_temperature_2021}. This prediction is consistent with the experimental data presented in Fig. \ref{fig:Fig1_aRuCl3_0Field_Alpha_Cones}(e).

Regarding the anisotropy of the attenuation, it was shown in \cite{ye_phonon_2020}, that the dominant contribution is due to the $E_2$ irreducible representation of the $C_{6v}$ point group. For the transverse mode, this results in an angular dependence of $ (1+\cos 6\theta_q)$ [see inset, Fig. \ref{fig:Fig1_aRuCl3_0Field_Alpha_Cones}(e)]. This leads to a maximum attenuation of the transverse mode for $\textbf{q}\parallel \textbf{a}$ and none for $\textbf{q}\parallel \textbf{b}$. In addition to the $E_2$ channel, however, contributions from a subdominant $A_1$ channel are also present, generating an angular-independent contribution to $\Delta\alpha(T)$. This is consistent with the observed angular variation of the slope in Fig.\ref{fig:Fig1_aRuCl3_0Field_Alpha_Cones} (e).  
Here, we also note that while our consideration is based on the layer group \textit{P}6\textit{mm} (layer group number 77), the lower symmetry of $\alpha$-RuCl$_3$ may further enhance the anisotropy of the sound attenuation coefficient along different crystallographic directions.

\begin{table}[tb]
	\centering
    \caption{Sound velocities and wave numbers for various acoustic modes and experimental settings. Here,
	$f$ is the frequency, $\textbf{q}$ the wave vector, $\textbf{u}$ the polarization of the acoustic wave, $v_{s}$ the sound velocity measured at 0.3 K, and $q$ the corresponding wave number. $v_s$  for both longitudinal (\cUS{11}) and transverse [$(\cUS{11} - \cUS{12})/{2}$] modes  agrees well with results from x-ray experiments from Li et al. \cite{li_giant_2021, li_divergence_2021} and Lebert et al. \cite{Lebert2022}.} 
	\label{tab:Schallgeschwindigkeiten}
	\begin{tabular*}{\columnwidth}{@{\extracolsep{\stretch{1}}}*{6}{lllll}@{}}
		\hline
        \hline
		\rule{0pt}{3ex} Mode 	& Geometry &
		$f$ & $v_{s}$		&  $q=\frac{2\pi f}{v}$\\
		&& [MHz]	& [m/s] & [nm$^{-1}$] \\		\hline 
		\cUS{33}	& $\textbf{q}\parallel \textbf{c}$;  $\textbf{u} \parallel \textbf{c}$ 
		& 106 & $2900 \pm 100$& 0.23 \\
		\cUS{44}	& $\textbf{q} \parallel \textbf{b}$;  $\textbf{u} \parallel \textbf{c}$ 
		& 27 & $1350 \pm 50$ & 0.126\\
		
		& Voigt &&& \\
		\cUS{11}	& $\textbf{q} \parallel \textbf{b}$;  $\textbf{u}\parallel \textbf{b}$ 
		& 40 & $3000 \pm 200$ 	& 0.084	\\
		$(\cUS{11} - \cUS{12})/{2}$ 
		& $\textbf{q} \parallel \textbf{b}$; $\textbf{u} \parallel \textbf{a}$ 	
		&  28 & $2400 \pm 100$	& 0.073	\\
		
		& Faraday &&& \\
		\cUS{11}	& $\textbf{q} \parallel \textbf{a}$;  $\textbf{u}\parallel \textbf{a}$ 
		&  22 & $3300 \pm 200$ & 0.042 \\
		$(\cUS{11} - \cUS{12})/{2}$
		& $\textbf{q} \parallel \textbf{a}$; $\textbf{u} \parallel \textbf{b}$ 
		& 26 & $2500\pm 100$ & 0.055	\\
		\hline
        \hline
	\end{tabular*}
	
\end{table}
  
Turning to the longitudinal $c_{11}$ mode, the sound velocities $v_{s}^{La}$ and $v_{s}^{Lb}$ from Table I for $\textbf{q}\parallel \textbf{a}$ and $\textbf{q}\parallel \textbf{b}$, respectively, are both larger than the Fermi velocity $v_F$, being in agreement with Ref. \cite{li_divergence_2021}. In that case, the kinematic constraints can only be satisfied in microscopic processes where a phonon decays into two fermions, both with positive energy -- dubbed pp-channel in Fig. \ref{fig:Fig1_aRuCl3_0Field_Alpha_Cones}(g). These processes happen even at zero temperature, if a phonon has enough energy to excite a pair of particles \cite{metavitsiadis_phonon_2020, ye_phonon_2020,feng_temperature_2021}. Consequently, $\Delta\alpha(T)$  depends on temperature only above the energy scale of the phonon. In that range, thermal occupation of the fermion states will block the scattering states due to the Pauli exclusion principle. This is consistent with the downturn of \cUS{11} in Fig. \ref{fig:Fig1_aRuCl3_0Field_Alpha_Cones}(e). We note, that asymptotically the Fermi function requires $\Delta\alpha(T) \propto T^{-1}$ for $T\gg v_F q$ \cite{li_divergence_2021}. 

Next, we comment on the relevance of the $\mathbb{Z}_2$ flux excitations. Specific-heat analysis \cite{Nasu2015, Metavitsiadis2017} shows that at zero magnetic field, complete proliferation of flux occurs in a very narrow temperature range of $T^\star \sim 0.01 \dots 0.03J < \unit[3]{K}$. Since this is clearly lower than \unit[10]{K}, i. e., the temperatures for which we discuss peculiarities of the sound attenuation  due to fractionalization
[Fig. \ref{fig:Fig1_aRuCl3_0Field_Alpha_Cones}(e)], flux excitations play a role. Here, two comments are due. First, numerical treatment \cite{metavitsiadis_phonon_2020, ye_phonon_2020,feng_temperature_2021} clarifies, that the emergent randomness due to static $\mathbb{Z}_2$ fluxes does not obliterate the presence of a characteristic energy scale $\omega_c$ ($\omega_c=v_F q$ in the absence of $\mathbb{Z}_2$ flux excitations), such that $\Delta\alpha(T)$ scales qualitatively similar to either ph- or pp-scattering for phonons above or below this scale, respectively (see also Fig. 4  in Ref. \cite{metavitsiadis_phonon_2020}). While $v_F$ of the flux-free sector is not a well-defined quantity at finite flux density, it is reasonable to assume that $\omega_c \sim v_F q$ remains true. Hence, we believe that our arguments based on the flux-free sector are qualitatively valid even in the temperature range of interest presented in Fig. \ref{fig:Fig1_aRuCl3_0Field_Alpha_Cones}(e). As a second comment, we stress that apart from the $A_1$ channel, finite flux density will additionally reduce the anisotropy of the sound attenuation since relaxing the  kinematic constraint on momenta allows for more scattering processes \cite{feng_temperature_2021}. For completeness, we mention that interactions beyond the pure Kitaev limit may further reduce the anisotropy in the pp-channel \cite{Susmita2023}.

The behavior of $\Delta\alpha(T)$ in a finite magnetic field is also quite remarkable. We obtain the corresponding data in Fig. \ref{fig:Fig1_aRuCl3_0Field_Alpha_Cones}(e), mostly at \unit[8]{T}, which is above the critical field $\mu_0 H_{c2}$ = \unit[6.9]{T}, suppressing the zero-field magnetic zigzag order in our sample \cite{wang_magnetic_2017}. For the $\cUS{11}$ mode with $\textbf{q} \parallel \textbf{a}$, data is available only at \unit[6]{T}. In the temperature range \unit[10--35]{K}, $\Delta\alpha(T)$ of the transverse modes for both $\textbf{q} \parallel \textbf{a}$ and $\textbf{q} \parallel \textbf{b}$ remains linear in $T$, consistent with the dominance of the ph-processes. However, the slope of the curve for $\textbf{q} \parallel \textbf{b}$ strongly increases compared to the zero-field data, while  it remains almost unchanged for the transverse mode with $\textbf{q} \parallel \textbf{a}$. So, in a finite magnetic field, the slopes of the $(\cUS{11}-\cUS{12})/2$ mode at $\textbf{q} \parallel \textbf{a}$ and $\textbf{q} \parallel \textbf{b}$ are becoming similar. This can be understood by the field-induced decrease of the flux-proliferation temperature $T^\star$ due to flux mobility \cite{Trebst2022}. At fixed $T$, this leads to a larger flux density at $\mu_0 H=\unit[8]{T}$ than at \unit[0]{T}, implying less anisotropy of $\Delta \alpha (T)$ for the transverse modes.

In the presence of a magnetic field, the sound attenuation of the longitudinal \cUS{11} mode requires more attention since the dominant contribution comes from the pp-processes.
First, in the flux-free pure Kitaev case, we expect a single-fermion gap of $\Delta\epsilon\sim \unit[10]{K}$, based on the perturbative parameter $\kappa/J\sim h^3/J^3$ \cite{kitaev_anyons_2006}. With $h=g\mu_B H$, one estimates $\kappa/J\sim 0.04$ at \unit[8]{T} \cite{Motome2019}.  The presence of this fermion gap, however, does not  strongly  affect the scattering of the phonons and, in particular, does not suppress the pp-processes, even though from Table \ref{tab:Schallgeschwindigkeiten} the phonon energies satisfy $\omega_q 
\ll \Delta\epsilon$. This  can be understood by the fact that the low-energy flux excitations in the presence of flux disorder \cite{tsh}
completely fill the field-induced gap  in the Majorana fermion's spectrum \cite{Metavitsiadis2022}. 
Next, we realize that the dominant contribution to the sound attenuation  $\Delta \alpha (T)$ for the longitudinal mode changes from the pp- to the ph-like for $\textbf{q} \parallel \textbf{b}$, while it remains pp-like for $\textbf{q}\parallel \textbf{a}$. This is consistent with the observation  that $v_{s}^{La}>v_{s}^{Lb}>v_F$ at zero field
 (see Table I) and with the enhanced low-$T$ softening of the elastic mode close to $H_{c1}$ and $H_{c2}$
 [shown in Fig. \ref{fig:aRuCl3_c11_kIIa_Examples_Phasediagram_Tdep_Hdep}(a) for the longitudinal \cUS{11} acoustic mode with $\textbf{q}\parallel  \textbf{a}$].
 From this, we speculate that a situation of $v_{s}^{La}>v_F>v_{s}^{Lb}$ arises in this field regime and leads to the change of the character in the temperature dependence of
 the  sound attenuation of the longitudinal  $c_{11}$ mode with $\textbf{q} \parallel \textbf{b}$.
 We also note that for the flux-free zero-field case, analysis of the pp-channel yields $\Delta \alpha (T) \sim q^3/T$ for $T\gg q v_F$ \cite{li_divergence_2021}.  Assuming that   these 
 estimates remain  qualitatively valid  at finite field and flux density and combining them with $[(q_{{\parallel} a})/(q_{{\parallel}b})]^3\sim 1/8$ from Table \ref{tab:Schallgeschwindigkeiten}, it is conceivable that attenuation phenomena are relatively weak for the $\cUS{11}$  mode with $\textbf{q}\parallel \textbf{a}$. 

Finally, we comment on the  low-temperature finite-field behavior of the sound attenuation. As we can see in Fig. \ref{fig:Fig1_aRuCl3_0Field_Alpha_Cones}(e), all curves show a non monotonous behavior at  temperatures below 10 K (this is less evident  for the $\cUS{11}$ mode with $\textbf{q} \parallel \textbf{b}$, since it happens at very low temperatures). 
To rationalize this observation, we recall that  there is 
a non negligible  amount of disorder in $\alpha$-RuCl$_3$,  which includes  vacancies, bond- and stacking-type of disorder \cite{Plumb2014,Sears2015}. 
Recent studies have shown that in the Kitaev spin liquid, disorder induces quasi-localized low-energy states, which at finite  field (6 or \unit[8]{T}), when dispersive fermionic modes are gapped, form in-gap low-energy bands \cite{Knolle2019,Kao2021,Kao2021localization,Vitor2022}.  These in-gap states provide an additional, predominantly pp-channel for phonon scattering, bringing about an increased sound attenuation at very low temperatures. 
However, once the temperature increases and  all these  in-gap states become populated, the attenuation exhibits a decrease  before starting to increase again due to the  phonon-scattering processes involving dispersive Majorana modes.

\subsection{$H$-$T$ phase diagram of $\alpha$-RuCl$_3$}
In Figs. \ref{fig:aRuCl3_c11_kIIa_Examples_Phasediagram_Tdep_Hdep}(a) and \ref{fig:aRuCl3_c11_kIIa_Examples_Phasediagram_Tdep_Hdep}(b), we show our results for the relative change of the sound velocity $\Delta v/v$ versus temperature and magnetic field, respectively, for the longitudinal mode \cUS{11} with $\textbf{q}\parallel \textbf{a}$ and $\textbf{H}\parallel\textbf{a}$.
Figure \ref{fig:aRuCl3_c11_kIIa_Examples_Phasediagram_Tdep_Hdep} clearly displays anomalies, marked by black arrows, both in the
 temperature (a) and magnetic-field (b) dependences of $\Delta v/v$. These anomalies signal phase transitions or crossover regimes in $\alpha$-RuCl$_3$. The anomaly at $T_N$ shifts with magnetic field and an additional step-like feature together with a shallow minimum appears at about \unit[1]{K}, for magnetic fields close to $H_{c2}$, [Fig. \ref{fig:aRuCl3_c11_kIIa_Examples_Phasediagram_Tdep_Hdep}(a)]. Remarkably, the elastic softening for $H$ close to $H_{c2}$ is at least twice as large as in zero field. Presumably, this is due to enhanced fluctuations in the quantum-critical regime. In fields $H$ $>$ $H_{c2}$, the velocity versus temperature exhibits a broad minimum which shifts to higher temperatures with increasing magnetic field [Fig. \ref{fig:aRuCl3_c11_kIIa_Examples_Phasediagram_Tdep_Hdep}(a)]. This minimum evidences a crossover regime in this range of temperatures and magnetic fields.

The field dependence of $\Delta v/v$ in Fig. \ref{fig:aRuCl3_c11_kIIa_Examples_Phasediagram_Tdep_Hdep}(b) shows two pronounced minima at $T=\unit[1]{K}$ (orange curve), one at $\mu_0 H_{c2}$ $\simeq$ \unit[6.9]{T} and the other, $\mu_0 H_{c1}$, just below the critical field (at around \unit[6]{T}). The anomaly at $H_{c2}$ shifts with temperature whereas $H_{c1}$ stays constant. With $H_{c2}$ reaching \unit[6]{T} close to \unit[5]{K}, both anomalies merge and for $T< T_N$ we observe a single anomaly in the field dependence. 
Moreover, we observe a change of curvature at small magnetic fields for temperatures below $T_N$. The large anomalies at 6 and \unit[6.9]{T} at \unit[1]{K} are related to the changes in the 3D magnetic structure between zigzag phases with different stacking orders \cite{balz_field_2021, balz_phasediagram_2019, schoenemann_thermal_2020} and the critical field $H_{c2}$, respectively. The low-field feature close to \unit[1]{T} might be due to a domain rearrangement in magnetic fields \cite{Sears2017}. Moreover, at higher temperatures, i. e., $T \geq \unit[7]{K}$, a broad minimum appears in the sound velocity [Fig. \ref{fig:aRuCl3_c11_kIIa_Examples_Phasediagram_Tdep_Hdep}(b)]. Its position correlates with the minimum observed in the temperature dependence of the sound velocity in this magnetic-field and temperature range [Fig. \ref{fig:aRuCl3_c11_kIIa_Examples_Phasediagram_Tdep_Hdep}(a)].

We summarize our observations by plotting the positions of all detected anomalies on an $H$-$T$ phase diagram in Fig. \ref{fig:aRuCl3_c11_kIIa_Examples_Phasediagram_Tdep_Hdep}(c).  Despite extensive studies \cite{baek_evidence_2017, hentrich_unusual_2018, balz_phasediagram_2019, schoenemann_thermal_2020, balz_field_2021} this is still under debate. In particular, while it is well known that magnetic fields applied along the ${\bf a}$ direction suppress the zigzag magnetic order in $\alpha$-RuCl$_3$, leading to a spin-liquid-like state, the precise nature of this state and whether or not it is adiabatically connected to the high-field paramagnetic phases remains an open issue despite extensive studies. 
The dashed line above \unit[8]{T} is consistent with the linearly increasing spin gap detected also by other methods \cite{PhysRevB.96.041405, Sears2017, hentrich_unusual_2018}. Remarkably, the sound-velocity minimum related to this crossover can be traced down nearly to $H_{c2}$.

\section{Summary and discussions}
We present the temperature and in-plane magnetic-field dependence of the relative velocity and attenuation of longitudinal and transverse sound waves in $\alpha$-RuCl$_3$. Strong spin-lattice coupling allows our ultrasound measurements to uncover remarkably rich information about the magnetic ground state and excitations of this material. We argue that the peculiar behavior of the sound attenuation in $\alpha$-RuCl$_3$ over a wide range of temperatures can be understood as a fingerprint of the fractionalized excitations of a Kitaev magnet with comparable Fermi and sound velocities: transverse sound modes with smaller velocities  scatter off from particle-hole excitations, inducing a characteristic $T$-linear damping, while longitudinal modes  with higher velocities  tap into the fermionic particle-particle continuum  at zero field. 
The $\mathbb{Z}_2$ flux excitations are relevant to lift part of the symmetry-required anisotropies of the attenuation, as well as to explain its high-field behavior. Our data suggest the striking scenario of a field-driven tuning of one of the longitudinal sound velocities through the Fermi velocity. Likely, these findings survive inclusion of further exchange interactions in the QSL phase \cite{Susmita2023}. 
Regarding the sound velocity, we identify a large set of transitions in temperature and in-plane magnetic-field dependences, and draw an $H$-$T$ phase diagram, consistent with previous work by other methods.
\begin{acknowledgements}
We thank Markus König at the Max Planck Institute for chemical Physics of Solids in Dresden for the technical support in the focused-ion-beam assisted polishing process.
We acknowledge support of the HLD at HZDR, member of the European Magnetic Field Laboratory (EMFL). Work of S.Z., W.B., A.H., and J.W. has been supported in part by the DFG through SFB 1143 (project-id 247310070). Work of A.H. and J.W. has been supported in part by the DFG trough excellence cluster $ct.qmat$ (EXC 2147, project-id 39085490).  W.B. acknowledges kind hospitality of the PSM, Dresden. The work of N.B.P. was supported by the U.S. Department of Energy, Office of Science, Basic Energy Sciences under Award No. DE-SC0018056.  N.B.P.  also acknowledges the hospitality and partial support  of the Technical University of Munich – Institute for Advanced
Study. M.Y.\ was supported by a grant from the Simons Foundation (216179, LB), and by the National Science Foundation under Grant No.\ NSF PHY-1748958.
The work of K.Y.C. at SKKU is supported by the National Research Foundation (NRF) of Korea (Grant No. 2020R1A5A1016518).
The work of V.T. was supported by the DFG through Transregional Research Collaboration TRR 80 (Augsburg, Munich, and Stuttgart) as well as by the project ANCD 20.80009.5007.19 (Moldova).
\end{acknowledgements}
\bibliography{bibliography.bib}

\begin{thebibliography}{73}%
\makeatletter
\providecommand \@ifxundefined [1]{%
 \@ifx{#1\undefined}
}%
\providecommand \@ifnum [1]{%
 \ifnum #1\expandafter \@firstoftwo
 \else \expandafter \@secondoftwo
 \fi
}%
\providecommand \@ifx [1]{%
 \ifx #1\expandafter \@firstoftwo
 \else \expandafter \@secondoftwo
 \fi
}%
\providecommand \natexlab [1]{#1}%
\providecommand \enquote  [1]{``#1''}%
\providecommand \bibnamefont  [1]{#1}%
\providecommand \bibfnamefont [1]{#1}%
\providecommand \citenamefont [1]{#1}%
\providecommand \href@noop [0]{\@secondoftwo}%
\providecommand \href [0]{\begingroup \@sanitize@url \@href}%
\providecommand \@href[1]{\@@startlink{#1}\@@href}%
\providecommand \@@href[1]{\endgroup#1\@@endlink}%
\providecommand \@sanitize@url [0]{\catcode `\\12\catcode `\$12\catcode
  `\&12\catcode `\#12\catcode `\^12\catcode `\_12\catcode `\%12\relax}%
\providecommand \@@startlink[1]{}%
\providecommand \@@endlink[0]{}%
\providecommand \url  [0]{\begingroup\@sanitize@url \@url }%
\providecommand \@url [1]{\endgroup\@href {#1}{\urlprefix }}%
\providecommand \urlprefix  [0]{URL }%
\providecommand \Eprint [0]{\href }%
\providecommand \doibase [0]{https://doi.org/}%
\providecommand \selectlanguage [0]{\@gobble}%
\providecommand \bibinfo  [0]{\@secondoftwo}%
\providecommand \bibfield  [0]{\@secondoftwo}%
\providecommand \translation [1]{[#1]}%
\providecommand \BibitemOpen [0]{}%
\providecommand \bibitemStop [0]{}%
\providecommand \bibitemNoStop [0]{.\EOS\space}%
\providecommand \EOS [0]{\spacefactor3000\relax}%
\providecommand \BibitemShut  [1]{\csname bibitem#1\endcsname}%
\let\auto@bib@innerbib\@empty
\bibitem [{\citenamefont {Kitaev}(2006)}]{kitaev_anyons_2006}%
  \BibitemOpen
  \bibfield  {author} {\bibinfo {author} {\bibfnamefont {A.}~\bibnamefont
  {Kitaev}},\ }\bibfield  {title} {\bibinfo {title} {\textit{{Anyons} in an
  {Exactly} {Solved} {Model} and {Beyond}}},\ }\href
  {https://doi.org/https://doi.org/10.1016/0022-3697(70)90163-0} {\bibfield
  {journal} {\bibinfo  {journal} {Ann. Phys.}\ }\textbf {\bibinfo {volume}
  {321}},\ \bibinfo {pages} {2} (\bibinfo {year} {2006})}\BibitemShut {NoStop}%
\bibitem [{\citenamefont {Baskaran}\ \emph {et~al.}(2007)\citenamefont
  {Baskaran}, \citenamefont {Mandal},\ and\ \citenamefont
  {Shankar}}]{baskaran_exact_2007}%
  \BibitemOpen
  \bibfield  {author} {\bibinfo {author} {\bibfnamefont {G.}~\bibnamefont
  {Baskaran}}, \bibinfo {author} {\bibfnamefont {S.}~\bibnamefont {Mandal}},\
  and\ \bibinfo {author} {\bibfnamefont {R.}~\bibnamefont {Shankar}},\
  }\bibfield  {title} {\bibinfo {title} {{\textit{Exact Results for Spin
  Dynamics and Fractionalization in the Kitaev Model}}},\ }\href
  {https://doi.org/10.1103/PhysRevLett.98.247201} {\bibfield  {journal}
  {\bibinfo  {journal} {Phys. Rev. Lett.}\ }\textbf {\bibinfo {volume} {98}},\
  \bibinfo {pages} {247201} (\bibinfo {year} {2007})}\BibitemShut {NoStop}%
\bibitem [{\citenamefont {Knolle}\ \emph
  {et~al.}(2014{\natexlab{a}})\citenamefont {Knolle}, \citenamefont
  {Kovrizhin}, \citenamefont {Chalker},\ and\ \citenamefont
  {Moessner}}]{knolle_dynamics_2014}%
  \BibitemOpen
  \bibfield  {author} {\bibinfo {author} {\bibfnamefont {J.}~\bibnamefont
  {Knolle}}, \bibinfo {author} {\bibfnamefont {D.~L.}\ \bibnamefont
  {Kovrizhin}}, \bibinfo {author} {\bibfnamefont {J.~T.}\ \bibnamefont
  {Chalker}},\ and\ \bibinfo {author} {\bibfnamefont {R.}~\bibnamefont
  {Moessner}},\ }\bibfield  {title} {\bibinfo {title} {\textit{{Dynamics of a
  Two-Dimensional Quantum Spin Liquid: Signatures of Emergent Majorana Fermions
  and Fluxes}}},\ }\href {https://doi.org/10.1103/PhysRevLett.112.207203}
  {\bibfield  {journal} {\bibinfo  {journal} {Phys. Rev. Lett.}\ }\textbf
  {\bibinfo {volume} {112}},\ \bibinfo {pages} {207203} (\bibinfo {year}
  {2014}{\natexlab{a}})}\BibitemShut {NoStop}%
\bibitem [{\citenamefont {Knolle}\ \emph {et~al.}(2015)\citenamefont {Knolle},
  \citenamefont {Kovrizhin}, \citenamefont {Chalker},\ and\ \citenamefont
  {Moessner}}]{Knolle2015}%
  \BibitemOpen
  \bibfield  {author} {\bibinfo {author} {\bibfnamefont {J.}~\bibnamefont
  {Knolle}}, \bibinfo {author} {\bibfnamefont {D.~L.}\ \bibnamefont
  {Kovrizhin}}, \bibinfo {author} {\bibfnamefont {J.~T.}\ \bibnamefont
  {Chalker}},\ and\ \bibinfo {author} {\bibfnamefont {R.}~\bibnamefont
  {Moessner}},\ }\bibfield  {title} {\bibinfo {title} {\textit{{Dynamics of
  Fractionalization in Quantum Spin Liquids}}},\ }\href
  {https://doi.org/10.1103/PhysRevB.92.115127} {\bibfield  {journal} {\bibinfo
  {journal} {Phys. Rev. B}\ }\textbf {\bibinfo {volume} {92}},\ \bibinfo
  {pages} {115127} (\bibinfo {year} {2015})}\BibitemShut {NoStop}%
\bibitem [{\citenamefont {Knolle}\ \emph
  {et~al.}(2014{\natexlab{b}})\citenamefont {Knolle}, \citenamefont {Chern},
  \citenamefont {Kovrizhin}, \citenamefont {Moessner},\ and\ \citenamefont
  {Perkins}}]{Knolle2014raman}%
  \BibitemOpen
  \bibfield  {author} {\bibinfo {author} {\bibfnamefont {J.}~\bibnamefont
  {Knolle}}, \bibinfo {author} {\bibfnamefont {G.-W.}\ \bibnamefont {Chern}},
  \bibinfo {author} {\bibfnamefont {D.~L.}\ \bibnamefont {Kovrizhin}}, \bibinfo
  {author} {\bibfnamefont {R.}~\bibnamefont {Moessner}},\ and\ \bibinfo
  {author} {\bibfnamefont {N.~B.}\ \bibnamefont {Perkins}},\ }\bibfield
  {title} {\bibinfo {title} {\textit{{Raman Scattering Signatures of Kitaev
  Spin Liquids in ${A}_{2}{\mathrm{IrO}}_{3}$ Iridates with $A=\mathrm{Na}$ or
  Li}}},\ }\href {https://doi.org/10.1103/PhysRevLett.113.187201} {\bibfield
  {journal} {\bibinfo  {journal} {Phys. Rev. Lett.}\ }\textbf {\bibinfo
  {volume} {113}},\ \bibinfo {pages} {187201} (\bibinfo {year}
  {2014}{\natexlab{b}})}\BibitemShut {NoStop}%
\bibitem [{\citenamefont {Nasu}\ \emph {et~al.}(2014)\citenamefont {Nasu},
  \citenamefont {Udagawa},\ and\ \citenamefont
  {Motome}}]{nasu2014vaporization}%
  \BibitemOpen
  \bibfield  {author} {\bibinfo {author} {\bibfnamefont {J.}~\bibnamefont
  {Nasu}}, \bibinfo {author} {\bibfnamefont {M.}~\bibnamefont {Udagawa}},\ and\
  \bibinfo {author} {\bibfnamefont {Y.}~\bibnamefont {Motome}},\ }\bibfield
  {title} {\bibinfo {title} {\textit{{Vaporization of Kitaev Spin Liquids}}},\
  }\href {https://journals.aps.org/prl/abstract/10.1103/PhysRevLett.113.197205}
  {\bibfield  {journal} {\bibinfo  {journal} {Phys. Rev. Lett.}\ }\textbf
  {\bibinfo {volume} {113}},\ \bibinfo {pages} {197205} (\bibinfo {year}
  {2014})}\BibitemShut {NoStop}%
\bibitem [{\citenamefont {Nasu}\ \emph {et~al.}(2016)\citenamefont {Nasu},
  \citenamefont {Knolle}, \citenamefont {Kovrizhin}, \citenamefont {Motome},\
  and\ \citenamefont {Moessner}}]{nasu2016fermionic}%
  \BibitemOpen
  \bibfield  {author} {\bibinfo {author} {\bibfnamefont {J.}~\bibnamefont
  {Nasu}}, \bibinfo {author} {\bibfnamefont {J.}~\bibnamefont {Knolle}},
  \bibinfo {author} {\bibfnamefont {D.~L.}\ \bibnamefont {Kovrizhin}}, \bibinfo
  {author} {\bibfnamefont {Y.}~\bibnamefont {Motome}},\ and\ \bibinfo {author}
  {\bibfnamefont {R.}~\bibnamefont {Moessner}},\ }\bibfield  {title} {\bibinfo
  {title} {\textit{{Fermionic Response from Fractionalization in an Insulating
  Two-Dimensional Magnet}}},\ }\href
  {https://www.nature.com/articles/nphys3809} {\bibfield  {journal} {\bibinfo
  {journal} {Nat. Phys.}\ }\textbf {\bibinfo {volume} {12}},\ \bibinfo {pages}
  {912} (\bibinfo {year} {2016})}\BibitemShut {NoStop}%
\bibitem [{\citenamefont {Hal\'asz}\ \emph {et~al.}(2016)\citenamefont
  {Hal\'asz}, \citenamefont {Perkins},\ and\ \citenamefont {van~den
  Brink}}]{halasz2016resonant}%
  \BibitemOpen
  \bibfield  {author} {\bibinfo {author} {\bibfnamefont {G.~B.}\ \bibnamefont
  {Hal\'asz}}, \bibinfo {author} {\bibfnamefont {N.~B.}\ \bibnamefont
  {Perkins}},\ and\ \bibinfo {author} {\bibfnamefont {J.}~\bibnamefont {van~den
  Brink}},\ }\bibfield  {title} {\bibinfo {title} {\textit{{Resonant Inelastic
  X-Ray Scattering Response of the Kitaev Honeycomb Model}}},\ }\href
  {https://doi.org/10.1103/PhysRevLett.117.127203} {\bibfield  {journal}
  {\bibinfo  {journal} {Phys. Rev. Lett.}\ }\textbf {\bibinfo {volume} {117}},\
  \bibinfo {pages} {127203} (\bibinfo {year} {2016})}\BibitemShut {NoStop}%
\bibitem [{\citenamefont {Hal\'asz}\ \emph {et~al.}(2019)\citenamefont
  {Hal\'asz}, \citenamefont {Kourtis}, \citenamefont {Knolle},\ and\
  \citenamefont {Perkins}}]{halasz2019observing}%
  \BibitemOpen
  \bibfield  {author} {\bibinfo {author} {\bibfnamefont {G.~B.}\ \bibnamefont
  {Hal\'asz}}, \bibinfo {author} {\bibfnamefont {S.}~\bibnamefont {Kourtis}},
  \bibinfo {author} {\bibfnamefont {J.}~\bibnamefont {Knolle}},\ and\ \bibinfo
  {author} {\bibfnamefont {N.~B.}\ \bibnamefont {Perkins}},\ }\bibfield
  {title} {\bibinfo {title} {\textit{{Observing Spin Fractionalization in the
  Kitaev Spin Liquid via Temperature Evolution of Indirect Resonant Inelastic
  X-Ray Scattering}}},\ }\href {https://doi.org/10.1103/PhysRevB.99.184417}
  {\bibfield  {journal} {\bibinfo  {journal} {Phys. Rev. B}\ }\textbf {\bibinfo
  {volume} {99}},\ \bibinfo {pages} {184417} (\bibinfo {year}
  {2019})}\BibitemShut {NoStop}%
\bibitem [{\citenamefont {Rousochatzakis}\ \emph {et~al.}(2019)\citenamefont
  {Rousochatzakis}, \citenamefont {Kourtis}, \citenamefont {Knolle},
  \citenamefont {Moessner},\ and\ \citenamefont
  {Perkins}}]{Rousochatzakis2019}%
  \BibitemOpen
  \bibfield  {author} {\bibinfo {author} {\bibfnamefont {I.}~\bibnamefont
  {Rousochatzakis}}, \bibinfo {author} {\bibfnamefont {S.}~\bibnamefont
  {Kourtis}}, \bibinfo {author} {\bibfnamefont {J.}~\bibnamefont {Knolle}},
  \bibinfo {author} {\bibfnamefont {R.}~\bibnamefont {Moessner}},\ and\
  \bibinfo {author} {\bibfnamefont {N.~B.}\ \bibnamefont {Perkins}},\
  }\bibfield  {title} {\bibinfo {title} {\textit{{Quantum Spin Liquid at Finite
  Temperature: Proximate Dynamics and Persistent Typicality}}},\ }\href
  {https://doi.org/10.1103/PhysRevB.100.045117} {\bibfield  {journal} {\bibinfo
   {journal} {Phys. Rev. B}\ }\textbf {\bibinfo {volume} {100}},\ \bibinfo
  {pages} {045117} (\bibinfo {year} {2019})}\BibitemShut {NoStop}%
\bibitem [{\citenamefont {Udagawa}\ \emph {et~al.}(2021)\citenamefont
  {Udagawa}, \citenamefont {Takayoshi},\ and\ \citenamefont
  {Oka}}]{udagawa2021}%
  \BibitemOpen
  \bibfield  {author} {\bibinfo {author} {\bibfnamefont {M.}~\bibnamefont
  {Udagawa}}, \bibinfo {author} {\bibfnamefont {S.}~\bibnamefont {Takayoshi}},\
  and\ \bibinfo {author} {\bibfnamefont {T.}~\bibnamefont {Oka}},\ }\bibfield
  {title} {\bibinfo {title} {\textit{{Scanning Tunneling Microscopy as a Single
  Majorana Detector of Kitaev's Chiral Spin Liquid}}},\ }\href
  {https://doi.org/10.1103/PhysRevLett.126.127201} {\bibfield  {journal}
  {\bibinfo  {journal} {Phys. Rev. Lett.}\ }\textbf {\bibinfo {volume} {126}},\
  \bibinfo {pages} {127201} (\bibinfo {year} {2021})}\BibitemShut {NoStop}%
\bibitem [{\citenamefont {Hermanns}\ \emph {et~al.}(2018)\citenamefont
  {Hermanns}, \citenamefont {Kimchi},\ and\ \citenamefont
  {Knolle}}]{Knolle2017}%
  \BibitemOpen
  \bibfield  {author} {\bibinfo {author} {\bibfnamefont {M.}~\bibnamefont
  {Hermanns}}, \bibinfo {author} {\bibfnamefont {I.}~\bibnamefont {Kimchi}},\
  and\ \bibinfo {author} {\bibfnamefont {J.}~\bibnamefont {Knolle}},\
  }\bibfield  {title} {\bibinfo {title} {\textit{{Physics of the Kitaev Model:
  Fractionalization, Dynamic Correlations, and Material Connections}}},\ }\href
  {https://doi.org/10.1146/annurev-conmatphys-033117-053934} {\bibfield
  {journal} {\bibinfo  {journal} {Annu. Rev. Condens. Matter Phys.}\ }\textbf
  {\bibinfo {volume} {9}},\ \bibinfo {pages} {17} (\bibinfo {year}
  {2018})}\BibitemShut {NoStop}%
\bibitem [{\citenamefont {Motome}\ and\ \citenamefont
  {Nasu}(2020)}]{Motome2019}%
  \BibitemOpen
  \bibfield  {author} {\bibinfo {author} {\bibfnamefont {Y.}~\bibnamefont
  {Motome}}\ and\ \bibinfo {author} {\bibfnamefont {J.}~\bibnamefont {Nasu}},\
  }\bibfield  {title} {\bibinfo {title} {\textit{{Hunting Majorana Fermions in
  Kitaev Magnets}}},\ }\href {https://doi.org/10.7566/JPSJ.89.012002}
  {\bibfield  {journal} {\bibinfo  {journal} {J. Phys. Soc. Jpn.}\ }\textbf
  {\bibinfo {volume} {89}},\ \bibinfo {pages} {012002} (\bibinfo {year}
  {2020})}\BibitemShut {NoStop}%
\bibitem [{\citenamefont {Trebst}\ and\ \citenamefont
  {Hickey}(2022)}]{Trebst2022}%
  \BibitemOpen
  \bibfield  {author} {\bibinfo {author} {\bibfnamefont {S.}~\bibnamefont
  {Trebst}}\ and\ \bibinfo {author} {\bibfnamefont {C.}~\bibnamefont
  {Hickey}},\ }\bibfield  {title} {\bibinfo {title} {\textit{{Kitaev
  Materials}}},\ }\href
  {https://doi.org/https://doi.org/10.1016/j.physrep.2021.11.003} {\bibfield
  {journal} {\bibinfo  {journal} {Phys. Rep.}\ }\textbf {\bibinfo {volume}
  {950}},\ \bibinfo {pages} {1} (\bibinfo {year} {2022})}\BibitemShut {NoStop}%
\bibitem [{\citenamefont {Zhou}\ and\ \citenamefont {Lee}(2011)}]{Plee2011}%
  \BibitemOpen
  \bibfield  {author} {\bibinfo {author} {\bibfnamefont {Y.}~\bibnamefont
  {Zhou}}\ and\ \bibinfo {author} {\bibfnamefont {P.~A.}\ \bibnamefont {Lee}},\
  }\bibfield  {title} {\bibinfo {title} {\textit{{Spinon Phonon Interaction and
  Ultrasonic Attenuation in Quantum Spin Liquids}}},\ }\href
  {https://doi.org/10.1103/PhysRevLett.106.056402} {\bibfield  {journal}
  {\bibinfo  {journal} {Phys. Rev. Lett.}\ }\textbf {\bibinfo {volume} {106}},\
  \bibinfo {pages} {056402} (\bibinfo {year} {2011})}\BibitemShut {NoStop}%
\bibitem [{\citenamefont {Serbyn}\ and\ \citenamefont {Lee}(2013)}]{Plee2013}%
  \BibitemOpen
  \bibfield  {author} {\bibinfo {author} {\bibfnamefont {M.}~\bibnamefont
  {Serbyn}}\ and\ \bibinfo {author} {\bibfnamefont {P.~A.}\ \bibnamefont
  {Lee}},\ }\bibfield  {title} {\bibinfo {title} {\textit{{Spinon-Phonon
  Interaction in Algebraic Spin Liquids}}},\ }\href
  {https://doi.org/10.1103/PhysRevB.87.174424} {\bibfield  {journal} {\bibinfo
  {journal} {Phys. Rev. B}\ }\textbf {\bibinfo {volume} {87}},\ \bibinfo
  {pages} {174424} (\bibinfo {year} {2013})}\BibitemShut {NoStop}%
\bibitem [{\citenamefont {Metavitsiadis}\ and\ \citenamefont
  {Brenig}(2020)}]{metavitsiadis_phonon_2020}%
  \BibitemOpen
  \bibfield  {author} {\bibinfo {author} {\bibfnamefont {A.}~\bibnamefont
  {Metavitsiadis}}\ and\ \bibinfo {author} {\bibfnamefont {W.}~\bibnamefont
  {Brenig}},\ }\bibfield  {title} {\bibinfo {title} {\textit{{Phonon
  Renormalization in the Kitaev Quantum Spin Liquid}}},\ }\href
  {https://doi.org/10.1103/PhysRevB.101.035103} {\bibfield  {journal} {\bibinfo
   {journal} {Phys. Rev. B}\ }\textbf {\bibinfo {volume} {101}},\ \bibinfo
  {pages} {035103} (\bibinfo {year} {2020})}\BibitemShut {NoStop}%
\bibitem [{\citenamefont {Ye}\ \emph {et~al.}(2020)\citenamefont {Ye},
  \citenamefont {Fernandes},\ and\ \citenamefont {Perkins}}]{ye_phonon_2020}%
  \BibitemOpen
  \bibfield  {author} {\bibinfo {author} {\bibfnamefont {M.}~\bibnamefont
  {Ye}}, \bibinfo {author} {\bibfnamefont {R.~M.}\ \bibnamefont {Fernandes}},\
  and\ \bibinfo {author} {\bibfnamefont {N.~B.}\ \bibnamefont {Perkins}},\
  }\bibfield  {title} {\bibinfo {title} {\textit{{Phonon Dynamics in the Kitaev
  Spin Liquid}}},\ }\href {https://doi.org/10.1103/PhysRevResearch.2.033180}
  {\bibfield  {journal} {\bibinfo  {journal} {Phys. Rev. Res.}\ }\textbf
  {\bibinfo {volume} {2}},\ \bibinfo {pages} {033180} (\bibinfo {year}
  {2020})}\BibitemShut {NoStop}%
\bibitem [{\citenamefont {Feng}\ \emph {et~al.}(2021)\citenamefont {Feng},
  \citenamefont {Ye},\ and\ \citenamefont {Perkins}}]{feng_temperature_2021}%
  \BibitemOpen
  \bibfield  {author} {\bibinfo {author} {\bibfnamefont {K.}~\bibnamefont
  {Feng}}, \bibinfo {author} {\bibfnamefont {M.}~\bibnamefont {Ye}},\ and\
  \bibinfo {author} {\bibfnamefont {N.~B.}\ \bibnamefont {Perkins}},\
  }\bibfield  {title} {\bibinfo {title} {\textit{{Temperature Evolution of the
  Phonon Dynamics in the Kitaev Spin Liquid}}},\ }\href
  {https://doi.org/10.1103/PhysRevB.103.214416} {\bibfield  {journal} {\bibinfo
   {journal} {Phys. Rev. B}\ }\textbf {\bibinfo {volume} {103}},\ \bibinfo
  {pages} {214416} (\bibinfo {year} {2021})}\BibitemShut {NoStop}%
\bibitem [{\citenamefont {Metavitsiadis}\ \emph {et~al.}(2022)\citenamefont
  {Metavitsiadis}, \citenamefont {Natori}, \citenamefont {Knolle},\ and\
  \citenamefont {Brenig}}]{Metavitsiadis2022}%
  \BibitemOpen
  \bibfield  {author} {\bibinfo {author} {\bibfnamefont {A.}~\bibnamefont
  {Metavitsiadis}}, \bibinfo {author} {\bibfnamefont {W.}~\bibnamefont
  {Natori}}, \bibinfo {author} {\bibfnamefont {J.}~\bibnamefont {Knolle}},\
  and\ \bibinfo {author} {\bibfnamefont {W.}~\bibnamefont {Brenig}},\
  }\bibfield  {title} {\bibinfo {title} {\textit{{Optical Phonons Coupled to a
  Kitaev Spin Liquid}}},\ }\href {https://doi.org/10.1103/PhysRevB.105.165151}
  {\bibfield  {journal} {\bibinfo  {journal} {Phys. Rev. B}\ }\textbf {\bibinfo
  {volume} {105}},\ \bibinfo {pages} {165151} (\bibinfo {year}
  {2022})}\BibitemShut {NoStop}%
\bibitem [{\citenamefont {Feng}\ \emph {et~al.}(2022)\citenamefont {Feng},
  \citenamefont {Swarup},\ and\ \citenamefont {Perkins}}]{Kexin2021}%
  \BibitemOpen
  \bibfield  {author} {\bibinfo {author} {\bibfnamefont {K.}~\bibnamefont
  {Feng}}, \bibinfo {author} {\bibfnamefont {S.}~\bibnamefont {Swarup}},\ and\
  \bibinfo {author} {\bibfnamefont {N.~B.}\ \bibnamefont {Perkins}},\
  }\bibfield  {title} {\bibinfo {title} {\textit{{Footprints of Kitaev Spin
  Liquid in the Fano Lineshape of Raman-Active Optical Phonons}}},\ }\href
  {https://doi.org/10.1103/PhysRevB.105.L121108} {\bibfield  {journal}
  {\bibinfo  {journal} {Phys. Rev. B}\ }\textbf {\bibinfo {volume} {105}},\
  \bibinfo {pages} {L121108} (\bibinfo {year} {2022})}\BibitemShut {NoStop}%
\bibitem [{\citenamefont {Plumb}\ \emph {et~al.}(2014)\citenamefont {Plumb},
  \citenamefont {Clancy}, \citenamefont {Sandilands}, \citenamefont {Shankar},
  \citenamefont {Hu}, \citenamefont {Burch}, \citenamefont {Kee},\ and\
  \citenamefont {Kim}}]{Plumb2014}%
  \BibitemOpen
  \bibfield  {author} {\bibinfo {author} {\bibfnamefont {K.~W.}\ \bibnamefont
  {Plumb}}, \bibinfo {author} {\bibfnamefont {J.~P.}\ \bibnamefont {Clancy}},
  \bibinfo {author} {\bibfnamefont {L.~J.}\ \bibnamefont {Sandilands}},
  \bibinfo {author} {\bibfnamefont {V.~V.}\ \bibnamefont {Shankar}}, \bibinfo
  {author} {\bibfnamefont {Y.~F.}\ \bibnamefont {Hu}}, \bibinfo {author}
  {\bibfnamefont {K.~S.}\ \bibnamefont {Burch}}, \bibinfo {author}
  {\bibfnamefont {H.-Y.}\ \bibnamefont {Kee}},\ and\ \bibinfo {author}
  {\bibfnamefont {Y.-J.}\ \bibnamefont {Kim}},\ }\bibfield  {title} {\bibinfo
  {title} {\textit{{$\ensuremath{\alpha}-{\mathrm{RuCl}}_{3}$: {A}
  {Spin}-{Orbit} {Assisted} {Mott} {Insulator} on a {Honeycomb} {Lattice}}}},\
  }\href {https://doi.org/10.1103/PhysRevB.90.041112} {\bibfield  {journal}
  {\bibinfo  {journal} {Phys. Rev. B}\ }\textbf {\bibinfo {volume} {90}},\
  \bibinfo {pages} {041112} (\bibinfo {year} {2014})}\BibitemShut {NoStop}%
\bibitem [{\citenamefont {Banerjee}\ \emph {et~al.}(2017)\citenamefont
  {Banerjee}, \citenamefont {Yan}, \citenamefont {Knolle}, \citenamefont
  {Bridges}, \citenamefont {Stone}, \citenamefont {Lumsden}, \citenamefont
  {Mandrus}, \citenamefont {Tennant}, \citenamefont {Moessner},\ and\
  \citenamefont {Nagler}}]{banerjee_neutron_2017}%
  \BibitemOpen
  \bibfield  {author} {\bibinfo {author} {\bibfnamefont {A.}~\bibnamefont
  {Banerjee}}, \bibinfo {author} {\bibfnamefont {J.}~\bibnamefont {Yan}},
  \bibinfo {author} {\bibfnamefont {J.}~\bibnamefont {Knolle}}, \bibinfo
  {author} {\bibfnamefont {C.~A.}\ \bibnamefont {Bridges}}, \bibinfo {author}
  {\bibfnamefont {M.~B.}\ \bibnamefont {Stone}}, \bibinfo {author}
  {\bibfnamefont {M.~D.}\ \bibnamefont {Lumsden}}, \bibinfo {author}
  {\bibfnamefont {D.~G.}\ \bibnamefont {Mandrus}}, \bibinfo {author}
  {\bibfnamefont {D.~A.}\ \bibnamefont {Tennant}}, \bibinfo {author}
  {\bibfnamefont {R.}~\bibnamefont {Moessner}},\ and\ \bibinfo {author}
  {\bibfnamefont {S.~E.}\ \bibnamefont {Nagler}},\ }\bibfield  {title}
  {\bibinfo {title} {\textit{{Neutron Scattering in the Proximate Quantum Spin
  Liquid \ensuremath{\alpha}-\ensuremath{\mathrm{RuCl}_3}}}},\ }\href
  {https://doi.org/10.1126/science.aah6015} {\bibfield  {journal} {\bibinfo
  {journal} {Science}\ }\textbf {\bibinfo {volume} {356}},\ \bibinfo {pages}
  {1055} (\bibinfo {year} {2017})}\BibitemShut {NoStop}%
\bibitem [{\citenamefont {Do}\ \emph {et~al.}(2017)\citenamefont {Do},
  \citenamefont {Park}, \citenamefont {Yoshitake}, \citenamefont {Nasu},
  \citenamefont {Motome}, \citenamefont {Kwon}, \citenamefont {Adroja},
  \citenamefont {Voneshen}, \citenamefont {Kim}, \citenamefont {Jang},
  \citenamefont {Park}, \citenamefont {Choi},\ and\ \citenamefont
  {Ji}}]{do_majorana_2017}%
  \BibitemOpen
  \bibfield  {author} {\bibinfo {author} {\bibfnamefont {S.-H.}\ \bibnamefont
  {Do}}, \bibinfo {author} {\bibfnamefont {S.-Y.}\ \bibnamefont {Park}},
  \bibinfo {author} {\bibfnamefont {J.}~\bibnamefont {Yoshitake}}, \bibinfo
  {author} {\bibfnamefont {J.}~\bibnamefont {Nasu}}, \bibinfo {author}
  {\bibfnamefont {Y.}~\bibnamefont {Motome}}, \bibinfo {author} {\bibfnamefont
  {Y.}~\bibnamefont {Kwon}}, \bibinfo {author} {\bibfnamefont {D.~T.}\
  \bibnamefont {Adroja}}, \bibinfo {author} {\bibfnamefont {D.~J.}\
  \bibnamefont {Voneshen}}, \bibinfo {author} {\bibfnamefont {K.}~\bibnamefont
  {Kim}}, \bibinfo {author} {\bibfnamefont {T.-H.}\ \bibnamefont {Jang}},
  \bibinfo {author} {\bibfnamefont {J.-H.}\ \bibnamefont {Park}}, \bibinfo
  {author} {\bibfnamefont {K.-Y.}\ \bibnamefont {Choi}},\ and\ \bibinfo
  {author} {\bibfnamefont {S.}~\bibnamefont {Ji}},\ }\bibfield  {title}
  {\bibinfo {title} {\textit{{Majorana Fermions in the {Kitaev} Quantum Spin
  System \ensuremath{\alpha}-\ensuremath{\mathrm{RuCl}_3}}}},\ }\href
  {https://doi.org/10.1038/nphys4264} {\bibfield  {journal} {\bibinfo
  {journal} {Nat. Phys.}\ }\textbf {\bibinfo {volume} {13}},\ \bibinfo {pages}
  {1079} (\bibinfo {year} {2017})}\BibitemShut {NoStop}%
\bibitem [{\citenamefont {Janša}\ \emph {et~al.}(2018)\citenamefont {Janša},
  \citenamefont {Zorko}, \citenamefont {Gomilšek}, \citenamefont {Pregelj},
  \citenamefont {Krämer}, \citenamefont {Biner}, \citenamefont {Biffin},
  \citenamefont {Rüegg},\ and\ \citenamefont
  {Klanjšek}}]{jansa_observation_2018}%
  \BibitemOpen
  \bibfield  {author} {\bibinfo {author} {\bibfnamefont {N.}~\bibnamefont
  {Janša}}, \bibinfo {author} {\bibfnamefont {A.}~\bibnamefont {Zorko}},
  \bibinfo {author} {\bibfnamefont {M.}~\bibnamefont {Gomilšek}}, \bibinfo
  {author} {\bibfnamefont {M.}~\bibnamefont {Pregelj}}, \bibinfo {author}
  {\bibfnamefont {K.~W.}\ \bibnamefont {Krämer}}, \bibinfo {author}
  {\bibfnamefont {D.}~\bibnamefont {Biner}}, \bibinfo {author} {\bibfnamefont
  {A.}~\bibnamefont {Biffin}}, \bibinfo {author} {\bibfnamefont
  {C.}~\bibnamefont {Rüegg}},\ and\ \bibinfo {author} {\bibfnamefont
  {M.}~\bibnamefont {Klanjšek}},\ }\bibfield  {title} {\bibinfo {title}
  {\textit{{Observation of two Types of Fractional Excitation in the {Kitaev}
  Honeycomb Magnet}}},\ }\href {https://doi.org/10.1038/s41567-018-0129-5}
  {\bibfield  {journal} {\bibinfo  {journal} {Nat. Phys.}\ }\textbf {\bibinfo
  {volume} {14}},\ \bibinfo {pages} {786} (\bibinfo {year} {2018})}\BibitemShut
  {NoStop}%
\bibitem [{\citenamefont {Takagi}\ \emph {et~al.}(2019)\citenamefont {Takagi},
  \citenamefont {Takayama}, \citenamefont {Jackeli}, \citenamefont
  {Khaliullin},\ and\ \citenamefont {Nagler}}]{takagi_concept_2019}%
  \BibitemOpen
  \bibfield  {author} {\bibinfo {author} {\bibfnamefont {H.}~\bibnamefont
  {Takagi}}, \bibinfo {author} {\bibfnamefont {T.}~\bibnamefont {Takayama}},
  \bibinfo {author} {\bibfnamefont {G.}~\bibnamefont {Jackeli}}, \bibinfo
  {author} {\bibfnamefont {G.}~\bibnamefont {Khaliullin}},\ and\ \bibinfo
  {author} {\bibfnamefont {S.~E.}\ \bibnamefont {Nagler}},\ }\bibfield  {title}
  {\bibinfo {title} {\textit{{Concept and Realization of {Kitaev} Quantum Spin
  Liquids}}},\ }\href {https://doi.org/10.1038/s42254-019-0038-2} {\bibfield
  {journal} {\bibinfo  {journal} {Nat. Rev. Phys.}\ }\textbf {\bibinfo {volume}
  {1}},\ \bibinfo {pages} {264} (\bibinfo {year} {2019})}\BibitemShut {NoStop}%
\bibitem [{\citenamefont {Sears}\ \emph {et~al.}(2015)\citenamefont {Sears},
  \citenamefont {Songvilay}, \citenamefont {Plumb}, \citenamefont {Clancy},
  \citenamefont {Qiu}, \citenamefont {Zhao}, \citenamefont {Parshall},\ and\
  \citenamefont {Kim}}]{Sears2015}%
  \BibitemOpen
  \bibfield  {author} {\bibinfo {author} {\bibfnamefont {J.~A.}\ \bibnamefont
  {Sears}}, \bibinfo {author} {\bibfnamefont {M.}~\bibnamefont {Songvilay}},
  \bibinfo {author} {\bibfnamefont {K.~W.}\ \bibnamefont {Plumb}}, \bibinfo
  {author} {\bibfnamefont {J.~P.}\ \bibnamefont {Clancy}}, \bibinfo {author}
  {\bibfnamefont {Y.}~\bibnamefont {Qiu}}, \bibinfo {author} {\bibfnamefont
  {Y.}~\bibnamefont {Zhao}}, \bibinfo {author} {\bibfnamefont {D.}~\bibnamefont
  {Parshall}},\ and\ \bibinfo {author} {\bibfnamefont {Y.-J.}\ \bibnamefont
  {Kim}},\ }\bibfield  {title} {\bibinfo {title} {\textit{{Magnetic Order in
  $\ensuremath{\alpha}-{\text{RuCl}}_{3}$: A Honeycomb-Lattice Quantum Magnet
  with Strong Spin-Orbit Coupling}}},\ }\href
  {https://doi.org/10.1103/PhysRevB.91.144420} {\bibfield  {journal} {\bibinfo
  {journal} {Phys. Rev. B}\ }\textbf {\bibinfo {volume} {91}},\ \bibinfo
  {pages} {144420} (\bibinfo {year} {2015})}\BibitemShut {NoStop}%
\bibitem [{\citenamefont {Balz}\ \emph {et~al.}(2021)\citenamefont {Balz},
  \citenamefont {Janssen}, \citenamefont {Lampen-Kelley}, \citenamefont
  {Banerjee}, \citenamefont {Liu}, \citenamefont {Yan}, \citenamefont
  {Mandrus}, \citenamefont {Vojta},\ and\ \citenamefont
  {Nagler}}]{balz_field_2021}%
  \BibitemOpen
  \bibfield  {author} {\bibinfo {author} {\bibfnamefont {C.}~\bibnamefont
  {Balz}}, \bibinfo {author} {\bibfnamefont {L.}~\bibnamefont {Janssen}},
  \bibinfo {author} {\bibfnamefont {P.}~\bibnamefont {Lampen-Kelley}}, \bibinfo
  {author} {\bibfnamefont {A.}~\bibnamefont {Banerjee}}, \bibinfo {author}
  {\bibfnamefont {Y.~H.}\ \bibnamefont {Liu}}, \bibinfo {author} {\bibfnamefont
  {J.-Q.}\ \bibnamefont {Yan}}, \bibinfo {author} {\bibfnamefont {D.~G.}\
  \bibnamefont {Mandrus}}, \bibinfo {author} {\bibfnamefont {M.}~\bibnamefont
  {Vojta}},\ and\ \bibinfo {author} {\bibfnamefont {S.~E.}\ \bibnamefont
  {Nagler}},\ }\bibfield  {title} {\bibinfo {title} {\textit{{Field-Induced
  Intermediate Ordered Phase and Anisotropic Interlayer Interactions in
  $\ensuremath{\alpha}\text{\ensuremath{-}}{\mathrm{RuCl}}_{3}$}}},\ }\href
  {https://doi.org/10.1103/PhysRevB.103.174417} {\bibfield  {journal} {\bibinfo
   {journal} {Phys. Rev. B}\ }\textbf {\bibinfo {volume} {103}},\ \bibinfo
  {pages} {174417} (\bibinfo {year} {2021})}\BibitemShut {NoStop}%
\bibitem [{\citenamefont {Suzuki}\ \emph {et~al.}(2021)\citenamefont {Suzuki},
  \citenamefont {Liu}, \citenamefont {Bertinshaw}, \citenamefont {Ueda},
  \citenamefont {Kim}, \citenamefont {Laha}, \citenamefont {Weber},
  \citenamefont {Yang}, \citenamefont {Wang}, \citenamefont {Takahashi},
  \citenamefont {Fürsich}, \citenamefont {Minola}, \citenamefont {Lotsch},
  \citenamefont {Kim}, \citenamefont {Yavaş}, \citenamefont {Daghofer},
  \citenamefont {Chaloupka}, \citenamefont {Khaliullin}, \citenamefont
  {Gretarsson},\ and\ \citenamefont {Keimer}}]{suzuki_proximate_2021}%
  \BibitemOpen
  \bibfield  {author} {\bibinfo {author} {\bibfnamefont {H.}~\bibnamefont
  {Suzuki}}, \bibinfo {author} {\bibfnamefont {H.}~\bibnamefont {Liu}},
  \bibinfo {author} {\bibfnamefont {J.}~\bibnamefont {Bertinshaw}}, \bibinfo
  {author} {\bibfnamefont {K.}~\bibnamefont {Ueda}}, \bibinfo {author}
  {\bibfnamefont {H.}~\bibnamefont {Kim}}, \bibinfo {author} {\bibfnamefont
  {S.}~\bibnamefont {Laha}}, \bibinfo {author} {\bibfnamefont {D.}~\bibnamefont
  {Weber}}, \bibinfo {author} {\bibfnamefont {Z.}~\bibnamefont {Yang}},
  \bibinfo {author} {\bibfnamefont {L.}~\bibnamefont {Wang}}, \bibinfo {author}
  {\bibfnamefont {H.}~\bibnamefont {Takahashi}}, \bibinfo {author}
  {\bibfnamefont {K.}~\bibnamefont {Fürsich}}, \bibinfo {author}
  {\bibfnamefont {M.}~\bibnamefont {Minola}}, \bibinfo {author} {\bibfnamefont
  {B.~V.}\ \bibnamefont {Lotsch}}, \bibinfo {author} {\bibfnamefont {B.~J.}\
  \bibnamefont {Kim}}, \bibinfo {author} {\bibfnamefont {H.}~\bibnamefont
  {Yavaş}}, \bibinfo {author} {\bibfnamefont {M.}~\bibnamefont {Daghofer}},
  \bibinfo {author} {\bibfnamefont {J.}~\bibnamefont {Chaloupka}}, \bibinfo
  {author} {\bibfnamefont {G.}~\bibnamefont {Khaliullin}}, \bibinfo {author}
  {\bibfnamefont {H.}~\bibnamefont {Gretarsson}},\ and\ \bibinfo {author}
  {\bibfnamefont {B.}~\bibnamefont {Keimer}},\ }\bibfield  {title} {\bibinfo
  {title} {\textit{{Proximate Ferromagnetic State in the {Kitaev} Model
  Material \ensuremath{\alpha}-{\ensuremath{\mathrm{RuCl}_3}}}}},\ }\href
  {https://doi.org/10.1038/s41467-021-24722-4} {\bibfield  {journal} {\bibinfo
  {journal} {Nat. Commun.}\ }\textbf {\bibinfo {volume} {12}},\ \bibinfo
  {pages} {4512} (\bibinfo {year} {2021})}\BibitemShut {NoStop}%
\bibitem [{\citenamefont {Wagner}\ \emph {et~al.}(2022)\citenamefont {Wagner},
  \citenamefont {Sahasrabudhe}, \citenamefont {Versteeg}, \citenamefont
  {Wysocki}, \citenamefont {Wang}, \citenamefont {Tsurkan}, \citenamefont
  {Loidl}, \citenamefont {Khomskii}, \citenamefont {Hedayat},\ and\
  \citenamefont {van Loosdrecht}}]{wagner_magneto-optical_2022}%
  \BibitemOpen
  \bibfield  {author} {\bibinfo {author} {\bibfnamefont {J.}~\bibnamefont
  {Wagner}}, \bibinfo {author} {\bibfnamefont {A.}~\bibnamefont
  {Sahasrabudhe}}, \bibinfo {author} {\bibfnamefont {R.~B.}\ \bibnamefont
  {Versteeg}}, \bibinfo {author} {\bibfnamefont {L.}~\bibnamefont {Wysocki}},
  \bibinfo {author} {\bibfnamefont {Z.}~\bibnamefont {Wang}}, \bibinfo {author}
  {\bibfnamefont {V.}~\bibnamefont {Tsurkan}}, \bibinfo {author} {\bibfnamefont
  {A.}~\bibnamefont {Loidl}}, \bibinfo {author} {\bibfnamefont {D.~I.}\
  \bibnamefont {Khomskii}}, \bibinfo {author} {\bibfnamefont {H.}~\bibnamefont
  {Hedayat}},\ and\ \bibinfo {author} {\bibfnamefont {P.~H.~M.}\ \bibnamefont
  {van Loosdrecht}},\ }\bibfield  {title} {\bibinfo {title}
  {\textit{{Magneto-Optical Study of Metamagnetic Transitions in the
  Antiferromagnetic Phase of
  \ensuremath{\alpha}-{\ensuremath{\mathrm{RuCl}_3}}}}},\ }\href
  {https://doi.org/10.1038/s41535-022-00434-w} {\bibfield  {journal} {\bibinfo
  {journal} {npj Quantum Mater.}\ }\textbf {\bibinfo {volume} {7}},\ \bibinfo
  {pages} {28} (\bibinfo {year} {2022})}\BibitemShut {NoStop}%
\bibitem [{\citenamefont {Bachus}\ \emph {et~al.}(2020)\citenamefont {Bachus},
  \citenamefont {Kaib}, \citenamefont {Tokiwa}, \citenamefont {Jesche},
  \citenamefont {Tsurkan}, \citenamefont {Loidl}, \citenamefont {Winter},
  \citenamefont {Tsirlin}, \citenamefont {Valentí},\ and\ \citenamefont
  {Gegenwart}}]{bachus_thermodynamic_2020}%
  \BibitemOpen
  \bibfield  {author} {\bibinfo {author} {\bibfnamefont {S.}~\bibnamefont
  {Bachus}}, \bibinfo {author} {\bibfnamefont {D.~A.~S.}\ \bibnamefont {Kaib}},
  \bibinfo {author} {\bibfnamefont {Y.}~\bibnamefont {Tokiwa}}, \bibinfo
  {author} {\bibfnamefont {A.}~\bibnamefont {Jesche}}, \bibinfo {author}
  {\bibfnamefont {V.}~\bibnamefont {Tsurkan}}, \bibinfo {author} {\bibfnamefont
  {A.}~\bibnamefont {Loidl}}, \bibinfo {author} {\bibfnamefont
  {S.}~\bibnamefont {Winter}}, \bibinfo {author} {\bibfnamefont
  {A.}~\bibnamefont {Tsirlin}}, \bibinfo {author} {\bibfnamefont
  {R.}~\bibnamefont {Valentí}},\ and\ \bibinfo {author} {\bibfnamefont
  {P.}~\bibnamefont {Gegenwart}},\ }\bibfield  {title} {\bibinfo {title}
  {\textit{{Thermodynamic {Perspective} on {Field}-{Induced} {Behavior} of
  \ensuremath{\alpha}-\ensuremath{\mathrm{RuCl}_3}}}},\ }\href
  {https://doi.org/10.1103/PhysRevLett.125.097203} {\bibfield  {journal}
  {\bibinfo  {journal} {Phys. Rev. Lett.}\ }\textbf {\bibinfo {volume} {125}},\
  \bibinfo {pages} {097203} (\bibinfo {year} {2020})}\BibitemShut {NoStop}%
\bibitem [{\citenamefont {Li}\ \emph {et~al.}(2021{\natexlab{a}})\citenamefont
  {Li}, \citenamefont {Zhang}, \citenamefont {Wang}, \citenamefont {Wu},
  \citenamefont {Gao}, \citenamefont {Qu}, \citenamefont {Liu}, \citenamefont
  {Gong},\ and\ \citenamefont {Li}}]{li_identification_2021}%
  \BibitemOpen
  \bibfield  {author} {\bibinfo {author} {\bibfnamefont {H.}~\bibnamefont
  {Li}}, \bibinfo {author} {\bibfnamefont {H.-K.}\ \bibnamefont {Zhang}},
  \bibinfo {author} {\bibfnamefont {J.}~\bibnamefont {Wang}}, \bibinfo {author}
  {\bibfnamefont {H.-Q.}\ \bibnamefont {Wu}}, \bibinfo {author} {\bibfnamefont
  {Y.}~\bibnamefont {Gao}}, \bibinfo {author} {\bibfnamefont {D.-W.}\
  \bibnamefont {Qu}}, \bibinfo {author} {\bibfnamefont {Z.-X.}\ \bibnamefont
  {Liu}}, \bibinfo {author} {\bibfnamefont {S.-S.}\ \bibnamefont {Gong}},\ and\
  \bibinfo {author} {\bibfnamefont {W.}~\bibnamefont {Li}},\ }\bibfield
  {title} {\bibinfo {title} {\textit{{Identification of Magnetic Interactions
  and High-Field Quantum Spin Liquid in
  \ensuremath{\alpha}-\ensuremath{\mathrm{RuCl}_3}}}},\ }\href
  {https://doi.org/10.1038/s41467-021-24257-8} {\bibfield  {journal} {\bibinfo
  {journal} {Nat. Commun.}\ }\textbf {\bibinfo {volume} {12}},\ \bibinfo
  {pages} {4007} (\bibinfo {year} {2021}{\natexlab{a}})}\BibitemShut {NoStop}%
\bibitem [{\citenamefont {Laurell}\ and\ \citenamefont
  {Okamoto}(2020)}]{laurell_dynamical_2020}%
  \BibitemOpen
  \bibfield  {author} {\bibinfo {author} {\bibfnamefont {P.}~\bibnamefont
  {Laurell}}\ and\ \bibinfo {author} {\bibfnamefont {S.}~\bibnamefont
  {Okamoto}},\ }\bibfield  {title} {\bibinfo {title} {\textit{{Dynamical and
  Thermal Magnetic Properties of the {Kitaev} Spin Liquid Candidate
  \ensuremath{\alpha}-\ensuremath{\mathrm{RuCl}_3}}}},\ }\href
  {https://doi.org/10.1038/s41535-019-0203-y} {\bibfield  {journal} {\bibinfo
  {journal} {npj Quantum Mater.}\ }\textbf {\bibinfo {volume} {5}},\ \bibinfo
  {pages} {2} (\bibinfo {year} {2020})}\BibitemShut {NoStop}%
\bibitem [{\citenamefont {Modic}\ \emph {et~al.}(2021)\citenamefont {Modic},
  \citenamefont {McDonald}, \citenamefont {Ruff}, \citenamefont {Bachmann},
  \citenamefont {Lai}, \citenamefont {Palmstrom}, \citenamefont {Graf},
  \citenamefont {Chan}, \citenamefont {Balakirev}, \citenamefont {Betts},
  \citenamefont {Boebinger}, \citenamefont {Schmidt}, \citenamefont {Lawler},
  \citenamefont {Sokolov}, \citenamefont {Moll}, \citenamefont {Ramshaw},\ and\
  \citenamefont {Shekhter}}]{modic_scale-invariant_2021}%
  \BibitemOpen
  \bibfield  {author} {\bibinfo {author} {\bibfnamefont {K.~A.}\ \bibnamefont
  {Modic}}, \bibinfo {author} {\bibfnamefont {R.~D.}\ \bibnamefont {McDonald}},
  \bibinfo {author} {\bibfnamefont {J.~P.~C.}\ \bibnamefont {Ruff}}, \bibinfo
  {author} {\bibfnamefont {M.~D.}\ \bibnamefont {Bachmann}}, \bibinfo {author}
  {\bibfnamefont {Y.}~\bibnamefont {Lai}}, \bibinfo {author} {\bibfnamefont
  {J.~C.}\ \bibnamefont {Palmstrom}}, \bibinfo {author} {\bibfnamefont
  {D.}~\bibnamefont {Graf}}, \bibinfo {author} {\bibfnamefont {M.~K.}\
  \bibnamefont {Chan}}, \bibinfo {author} {\bibfnamefont {F.~F.}\ \bibnamefont
  {Balakirev}}, \bibinfo {author} {\bibfnamefont {J.~B.}\ \bibnamefont
  {Betts}}, \bibinfo {author} {\bibfnamefont {G.~S.}\ \bibnamefont
  {Boebinger}}, \bibinfo {author} {\bibfnamefont {M.}~\bibnamefont {Schmidt}},
  \bibinfo {author} {\bibfnamefont {M.~J.}\ \bibnamefont {Lawler}}, \bibinfo
  {author} {\bibfnamefont {D.~A.}\ \bibnamefont {Sokolov}}, \bibinfo {author}
  {\bibfnamefont {P.~J.~W.}\ \bibnamefont {Moll}}, \bibinfo {author}
  {\bibfnamefont {B.~J.}\ \bibnamefont {Ramshaw}},\ and\ \bibinfo {author}
  {\bibfnamefont {A.}~\bibnamefont {Shekhter}},\ }\bibfield  {title} {\bibinfo
  {title} {\textit{{Scale-Invariant Magnetic Anisotropy in
  $\ensuremath{\alpha}\text{\ensuremath{-}}{\mathrm{RuCl}}_{3}$ at High
  Magnetic Fields}}},\ }\href {https://doi.org/10.1038/s41567-020-1028-0}
  {\bibfield  {journal} {\bibinfo  {journal} {Nat. Phys.}\ }\textbf {\bibinfo
  {volume} {17}},\ \bibinfo {pages} {240} (\bibinfo {year} {2021})}\BibitemShut
  {NoStop}%
\bibitem [{bae()}]{baek_evidence_2017}%
  \BibitemOpen
  \href@noop {} {\bibinfo  {journal} {S.-H. Baek, S.-H. Do, K.-Y. Choi, Y. S.
  Kwon, A. U. B. Wolter, S. Nishimoto, J. van den Brink, and B. B\"uchner,
  \textit{Evidence for a Field-Induced Quantum Spin Liquid in
  $\ensuremath{\alpha}\text{\ensuremath{-}}{\mathrm{RuCl}}_{3}$}, Phys. Rev.
  Lett. {\bf 119}, 037201 (2017)}\ }\BibitemShut {NoStop}%
\bibitem [{\citenamefont {Wolter}\ and\ \citenamefont
  {Hess}(2022)}]{wolter_spin_2022}%
  \BibitemOpen
\bibfield  {journal} {  }\bibfield  {author} {\bibinfo {author} {\bibfnamefont
  {A.~U.~B.}\ \bibnamefont {Wolter}}\ and\ \bibinfo {author} {\bibfnamefont
  {C.}~\bibnamefont {Hess}},\ }\bibfield  {title} {\bibinfo {title}
  {\textit{{Spin Liquid Evidence at the Edge and in Bulk}}},\ }\href
  {https://doi.org/10.1038/s41567-022-01521-2} {\bibfield  {journal} {\bibinfo
  {journal} {Nat. Phys.}\ }\textbf {\bibinfo {volume} {18}},\ \bibinfo {pages}
  {378} (\bibinfo {year} {2022})}\BibitemShut {NoStop}%
\bibitem [{\citenamefont {Kasahara}\ \emph {et~al.}(2018)\citenamefont
  {Kasahara}, \citenamefont {Ohnishi}, \citenamefont {Mizukami}, \citenamefont
  {Tanaka}, \citenamefont {Ma}, \citenamefont {Sugii}, \citenamefont {Kurita},
  \citenamefont {Tanaka}, \citenamefont {Nasu}, \citenamefont {Motome},
  \citenamefont {Shibauchi},\ and\ \citenamefont
  {Matsuda}}]{kasahara_majorana_2018}%
  \BibitemOpen
  \bibfield  {author} {\bibinfo {author} {\bibfnamefont {Y.}~\bibnamefont
  {Kasahara}}, \bibinfo {author} {\bibfnamefont {T.}~\bibnamefont {Ohnishi}},
  \bibinfo {author} {\bibfnamefont {Y.}~\bibnamefont {Mizukami}}, \bibinfo
  {author} {\bibfnamefont {O.}~\bibnamefont {Tanaka}}, \bibinfo {author}
  {\bibfnamefont {S.}~\bibnamefont {Ma}}, \bibinfo {author} {\bibfnamefont
  {K.}~\bibnamefont {Sugii}}, \bibinfo {author} {\bibfnamefont
  {N.}~\bibnamefont {Kurita}}, \bibinfo {author} {\bibfnamefont
  {H.}~\bibnamefont {Tanaka}}, \bibinfo {author} {\bibfnamefont
  {J.}~\bibnamefont {Nasu}}, \bibinfo {author} {\bibfnamefont {Y.}~\bibnamefont
  {Motome}}, \bibinfo {author} {\bibfnamefont {T.}~\bibnamefont {Shibauchi}},\
  and\ \bibinfo {author} {\bibfnamefont {Y.}~\bibnamefont {Matsuda}},\
  }\bibfield  {title} {\bibinfo {title} {\textit{{Majorana Quantization and
  Half-Integer Thermal Quantum {Hall} Effect in a {Kitaev} Spin Liquid}}},\
  }\href {https://doi.org/10.1038/s41586-018-0274-0} {\bibfield  {journal}
  {\bibinfo  {journal} {Nature}\ }\textbf {\bibinfo {volume} {559}},\ \bibinfo
  {pages} {227} (\bibinfo {year} {2018})}\BibitemShut {NoStop}%
\bibitem [{\citenamefont {Tanaka}\ \emph {et~al.}(2022)\citenamefont {Tanaka},
  \citenamefont {Mizukami}, \citenamefont {Harasawa}, \citenamefont
  {Hashimoto}, \citenamefont {Hwang}, \citenamefont {Kurita}, \citenamefont
  {Tanaka}, \citenamefont {Fujimoto}, \citenamefont {Matsuda}, \citenamefont
  {Moon},\ and\ \citenamefont {Shibauchi}}]{tanaka_thermodynamic_2022}%
  \BibitemOpen
  \bibfield  {author} {\bibinfo {author} {\bibfnamefont {O.}~\bibnamefont
  {Tanaka}}, \bibinfo {author} {\bibfnamefont {Y.}~\bibnamefont {Mizukami}},
  \bibinfo {author} {\bibfnamefont {R.}~\bibnamefont {Harasawa}}, \bibinfo
  {author} {\bibfnamefont {K.}~\bibnamefont {Hashimoto}}, \bibinfo {author}
  {\bibfnamefont {K.}~\bibnamefont {Hwang}}, \bibinfo {author} {\bibfnamefont
  {N.}~\bibnamefont {Kurita}}, \bibinfo {author} {\bibfnamefont
  {H.}~\bibnamefont {Tanaka}}, \bibinfo {author} {\bibfnamefont
  {S.}~\bibnamefont {Fujimoto}}, \bibinfo {author} {\bibfnamefont
  {Y.}~\bibnamefont {Matsuda}}, \bibinfo {author} {\bibfnamefont {E.-G.}\
  \bibnamefont {Moon}},\ and\ \bibinfo {author} {\bibfnamefont
  {T.}~\bibnamefont {Shibauchi}},\ }\bibfield  {title} {\bibinfo {title}
  {\textit{{Thermodynamic Evidence for a Field-Angle-Dependent {Majorana} Gap
  in a {Kitaev} Spin Liquid}}},\ }\href
  {https://doi.org/10.1038/s41567-021-01488-6} {\bibfield  {journal} {\bibinfo
  {journal} {Nat. Phys.}\ }\textbf {\bibinfo {volume} {18}},\ \bibinfo {pages}
  {429} (\bibinfo {year} {2022})}\BibitemShut {NoStop}%
\bibitem [{\citenamefont {Yokoi}\ \emph {et~al.}(2021)\citenamefont {Yokoi},
  \citenamefont {Ma}, \citenamefont {Kasahara}, \citenamefont {Kasahara},
  \citenamefont {Shibauchi}, \citenamefont {Kurita}, \citenamefont {Tanaka},
  \citenamefont {Nasu}, \citenamefont {Motome}, \citenamefont {Hickey},
  \citenamefont {Trebst},\ and\ \citenamefont
  {Matsuda}}]{yokoi_half-integer_2021}%
  \BibitemOpen
  \bibfield  {author} {\bibinfo {author} {\bibfnamefont {T.}~\bibnamefont
  {Yokoi}}, \bibinfo {author} {\bibfnamefont {S.}~\bibnamefont {Ma}}, \bibinfo
  {author} {\bibfnamefont {Y.}~\bibnamefont {Kasahara}}, \bibinfo {author}
  {\bibfnamefont {S.}~\bibnamefont {Kasahara}}, \bibinfo {author}
  {\bibfnamefont {T.}~\bibnamefont {Shibauchi}}, \bibinfo {author}
  {\bibfnamefont {N.}~\bibnamefont {Kurita}}, \bibinfo {author} {\bibfnamefont
  {H.}~\bibnamefont {Tanaka}}, \bibinfo {author} {\bibfnamefont
  {J.}~\bibnamefont {Nasu}}, \bibinfo {author} {\bibfnamefont {Y.}~\bibnamefont
  {Motome}}, \bibinfo {author} {\bibfnamefont {C.}~\bibnamefont {Hickey}},
  \bibinfo {author} {\bibfnamefont {S.}~\bibnamefont {Trebst}},\ and\ \bibinfo
  {author} {\bibfnamefont {Y.}~\bibnamefont {Matsuda}},\ }\bibfield  {title}
  {\bibinfo {title} {\textit{{Half-Integer Quantized Anomalous Thermal {Hall}
  Effect in the {Kitaev} Material Candidate
  \ensuremath{\alpha}-\ensuremath{\mathrm{RuCl}_3}}}},\ }\href
  {https://doi.org/10.1126/science.aay5551} {\bibfield  {journal} {\bibinfo
  {journal} {Science}\ }\textbf {\bibinfo {volume} {373}},\ \bibinfo {pages}
  {568} (\bibinfo {year} {2021})}\BibitemShut {NoStop}%
\bibitem [{\citenamefont {Bachus}\ \emph {et~al.}(2021)\citenamefont {Bachus},
  \citenamefont {Kaib}, \citenamefont {Jesche}, \citenamefont {Tsurkan},
  \citenamefont {Loidl}, \citenamefont {Winter}, \citenamefont {Tsirlin},
  \citenamefont {Valent\'{\i}},\ and\ \citenamefont
  {Gegenwart}}]{bachus_angle_2021}%
  \BibitemOpen
  \bibfield  {author} {\bibinfo {author} {\bibfnamefont {S.}~\bibnamefont
  {Bachus}}, \bibinfo {author} {\bibfnamefont {D.~A.~S.}\ \bibnamefont {Kaib}},
  \bibinfo {author} {\bibfnamefont {A.}~\bibnamefont {Jesche}}, \bibinfo
  {author} {\bibfnamefont {V.}~\bibnamefont {Tsurkan}}, \bibinfo {author}
  {\bibfnamefont {A.}~\bibnamefont {Loidl}}, \bibinfo {author} {\bibfnamefont
  {S.~M.}\ \bibnamefont {Winter}}, \bibinfo {author} {\bibfnamefont {A.~A.}\
  \bibnamefont {Tsirlin}}, \bibinfo {author} {\bibfnamefont {R.}~\bibnamefont
  {Valent\'{\i}}},\ and\ \bibinfo {author} {\bibfnamefont {P.}~\bibnamefont
  {Gegenwart}},\ }\bibfield  {title} {\bibinfo {title}
  {\textit{{Angle-Dependent Thermodynamics of
  $\ensuremath{\alpha}\text{\ensuremath{-}}\mathrm{Ru}{\mathrm{Cl}}_{3}$}}},\
  }\href {https://doi.org/10.1103/PhysRevB.103.054440} {\bibfield  {journal}
  {\bibinfo  {journal} {Phys. Rev. B}\ }\textbf {\bibinfo {volume} {103}},\
  \bibinfo {pages} {054440} (\bibinfo {year} {2021})}\BibitemShut {NoStop}%
\bibitem [{\citenamefont {Ye}\ \emph {et~al.}(2018)\citenamefont {Ye},
  \citenamefont {Hal{\'a}sz}, \citenamefont {Savary},\ and\ \citenamefont
  {Balents}}]{ye2018quantization}%
  \BibitemOpen
  \bibfield  {author} {\bibinfo {author} {\bibfnamefont {M.}~\bibnamefont
  {Ye}}, \bibinfo {author} {\bibfnamefont {G.~B.}\ \bibnamefont {Hal{\'a}sz}},
  \bibinfo {author} {\bibfnamefont {L.}~\bibnamefont {Savary}},\ and\ \bibinfo
  {author} {\bibfnamefont {L.}~\bibnamefont {Balents}},\ }\bibfield  {title}
  {\bibinfo {title} {\textit{{Quantization of the Thermal Hall Conductivity at
  Small Hall Angles}}},\ }\href@noop {} {\bibfield  {journal} {\bibinfo
  {journal} {Phys. Rev. Lett.}\ }\textbf {\bibinfo {volume} {121}},\ \bibinfo
  {pages} {147201} (\bibinfo {year} {2018})}\BibitemShut {NoStop}%
\bibitem [{\citenamefont {Vinkler-Aviv}\ and\ \citenamefont
  {Rosch}(2018)}]{vinkler-aviv_approximately_2018}%
  \BibitemOpen
  \bibfield  {author} {\bibinfo {author} {\bibfnamefont {Y.}~\bibnamefont
  {Vinkler-Aviv}}\ and\ \bibinfo {author} {\bibfnamefont {A.}~\bibnamefont
  {Rosch}},\ }\bibfield  {title} {\bibinfo {title} {\textit{{Approximately
  Quantized Thermal Hall Effect of Chiral Liquids Coupled to Phonons}}},\
  }\href {https://doi.org/10.1103/PhysRevX.8.031032} {\bibfield  {journal}
  {\bibinfo  {journal} {Phys. Rev. X}\ }\textbf {\bibinfo {volume} {8}},\
  \bibinfo {pages} {031032} (\bibinfo {year} {2018})}\BibitemShut {NoStop}%
\bibitem [{\citenamefont {Kocsis}\ \emph {et~al.}(2022)\citenamefont {Kocsis},
  \citenamefont {Kaib}, \citenamefont {Riedl}, \citenamefont {Gass},
  \citenamefont {Lampen-Kelley}, \citenamefont {Mandrus}, \citenamefont
  {Nagler}, \citenamefont {P\'erez}, \citenamefont {Nielsch}, \citenamefont
  {B\"uchner}, \citenamefont {Wolter},\ and\ \citenamefont
  {Valent\'{\i}}}]{kocsis_magnetoelastic_2022}%
  \BibitemOpen
  \bibfield  {author} {\bibinfo {author} {\bibfnamefont {V.}~\bibnamefont
  {Kocsis}}, \bibinfo {author} {\bibfnamefont {D.~A.~S.}\ \bibnamefont {Kaib}},
  \bibinfo {author} {\bibfnamefont {K.}~\bibnamefont {Riedl}}, \bibinfo
  {author} {\bibfnamefont {S.}~\bibnamefont {Gass}}, \bibinfo {author}
  {\bibfnamefont {P.}~\bibnamefont {Lampen-Kelley}}, \bibinfo {author}
  {\bibfnamefont {D.~G.}\ \bibnamefont {Mandrus}}, \bibinfo {author}
  {\bibfnamefont {S.~E.}\ \bibnamefont {Nagler}}, \bibinfo {author}
  {\bibfnamefont {N.}~\bibnamefont {P\'erez}}, \bibinfo {author} {\bibfnamefont
  {K.}~\bibnamefont {Nielsch}}, \bibinfo {author} {\bibfnamefont
  {B.}~\bibnamefont {B\"uchner}}, \bibinfo {author} {\bibfnamefont {A.~U.~B.}\
  \bibnamefont {Wolter}},\ and\ \bibinfo {author} {\bibfnamefont
  {R.}~\bibnamefont {Valent\'{\i}}},\ }\bibfield  {title} {\bibinfo {title}
  {\textit{{Magnetoelastic Coupling Anisotropy in the Kitaev Material
  $\ensuremath{\alpha}\text{\ensuremath{-}}\mathrm{Ru}{\mathrm{Cl}}_{3}$}}},\
  }\href {https://doi.org/10.1103/PhysRevB.105.094410} {\bibfield  {journal}
  {\bibinfo  {journal} {Phys. Rev. B}\ }\textbf {\bibinfo {volume} {105}},\
  \bibinfo {pages} {094410} (\bibinfo {year} {2022})}\BibitemShut {NoStop}%
\bibitem [{\citenamefont {Kaib}\ \emph {et~al.}(2021)\citenamefont {Kaib},
  \citenamefont {Biswas}, \citenamefont {Riedl}, \citenamefont {Winter},\ and\
  \citenamefont {Valent\'{\i}}}]{kaib_magnetoelastic_2021}%
  \BibitemOpen
  \bibfield  {author} {\bibinfo {author} {\bibfnamefont {D.~A.~S.}\
  \bibnamefont {Kaib}}, \bibinfo {author} {\bibfnamefont {S.}~\bibnamefont
  {Biswas}}, \bibinfo {author} {\bibfnamefont {K.}~\bibnamefont {Riedl}},
  \bibinfo {author} {\bibfnamefont {S.~M.}\ \bibnamefont {Winter}},\ and\
  \bibinfo {author} {\bibfnamefont {R.}~\bibnamefont {Valent\'{\i}}},\
  }\bibfield  {title} {\bibinfo {title} {\textit{{Magnetoelastic Coupling and
  Effects of Uniaxial Strain in
  $\ensuremath{\alpha}\ensuremath{-}{\mathrm{RuCl}}_{3}$ from First
  Principles}}},\ }\href {https://doi.org/10.1103/PhysRevB.103.L140402}
  {\bibfield  {journal} {\bibinfo  {journal} {Phys. Rev. B}\ }\textbf {\bibinfo
  {volume} {103}},\ \bibinfo {pages} {L140402} (\bibinfo {year}
  {2021})}\BibitemShut {NoStop}%
\bibitem [{\citenamefont {Sch\"onemann}\ \emph {et~al.}(2020)\citenamefont
  {Sch\"onemann}, \citenamefont {Imajo}, \citenamefont {Weickert},
  \citenamefont {Yan}, \citenamefont {Mandrus}, \citenamefont {Takano},
  \citenamefont {Brosha}, \citenamefont {Rosa}, \citenamefont {Nagler},
  \citenamefont {Kindo},\ and\ \citenamefont
  {Jaime}}]{schoenemann_thermal_2020}%
  \BibitemOpen
  \bibfield  {author} {\bibinfo {author} {\bibfnamefont {R.}~\bibnamefont
  {Sch\"onemann}}, \bibinfo {author} {\bibfnamefont {S.}~\bibnamefont {Imajo}},
  \bibinfo {author} {\bibfnamefont {F.}~\bibnamefont {Weickert}}, \bibinfo
  {author} {\bibfnamefont {J.}~\bibnamefont {Yan}}, \bibinfo {author}
  {\bibfnamefont {D.~G.}\ \bibnamefont {Mandrus}}, \bibinfo {author}
  {\bibfnamefont {Y.}~\bibnamefont {Takano}}, \bibinfo {author} {\bibfnamefont
  {E.~L.}\ \bibnamefont {Brosha}}, \bibinfo {author} {\bibfnamefont {P.~F.~S.}\
  \bibnamefont {Rosa}}, \bibinfo {author} {\bibfnamefont {S.~E.}\ \bibnamefont
  {Nagler}}, \bibinfo {author} {\bibfnamefont {K.}~\bibnamefont {Kindo}},\ and\
  \bibinfo {author} {\bibfnamefont {M.}~\bibnamefont {Jaime}},\ }\bibfield
  {title} {\bibinfo {title} {\textit{{Thermal and Magnetoelastic Properties of
  $\ensuremath{\alpha}\ensuremath{-}{\mathrm{RuCl}}_{3}$ in the Field-Induced
  Low-Temperature States}}},\ }\href
  {https://doi.org/10.1103/PhysRevB.102.214432} {\bibfield  {journal} {\bibinfo
   {journal} {Phys. Rev. B}\ }\textbf {\bibinfo {volume} {102}},\ \bibinfo
  {pages} {214432} (\bibinfo {year} {2020})}\BibitemShut {NoStop}%
\bibitem [{Mu2()}]{Mu2022}%
  \BibitemOpen
  \href@noop {} {\bibinfo  {journal} {S. Mu, K. D. Dixit, X. Wang, D. L.
  Abernathy, H. Cao, S. E. Nagler, J. Yan, P. Lampen-Kelley, D. Mandrus, C. A.
  Polanco, L. Liang, G. B. Hal\'asz, Y. Cheng, A. Banerjee, and T. Berlijn,
  \textit{Role of the Third Dimension in Searching for Majorana Fermions in
  $\ensuremath{\alpha}\text{\ensuremath{-}}{\mathrm{RuCl}}_{3}$ via Phonons},
  Phys. Rev. Res. {\bf 4}, 013067 (2022)}\ }\BibitemShut {NoStop}%
\bibitem [{\citenamefont {Hentrich}\ \emph {et~al.}(2018)\citenamefont
  {Hentrich}, \citenamefont {Wolter}, \citenamefont {Zotos}, \citenamefont
  {Brenig}, \citenamefont {Nowak}, \citenamefont {Isaeva}, \citenamefont
  {Doert}, \citenamefont {Banerjee}, \citenamefont {Lampen-Kelley},
  \citenamefont {Mandrus}, \citenamefont {Nagler}, \citenamefont {Sears},
  \citenamefont {Kim}, \citenamefont {B\"uchner},\ and\ \citenamefont
  {Hess}}]{hentrich_unusual_2018}%
  \BibitemOpen
\bibfield  {journal} {  }\bibfield  {author} {\bibinfo {author} {\bibfnamefont
  {R.}~\bibnamefont {Hentrich}}, \bibinfo {author} {\bibfnamefont {A.~U.~B.}\
  \bibnamefont {Wolter}}, \bibinfo {author} {\bibfnamefont {X.}~\bibnamefont
  {Zotos}}, \bibinfo {author} {\bibfnamefont {W.}~\bibnamefont {Brenig}},
  \bibinfo {author} {\bibfnamefont {D.}~\bibnamefont {Nowak}}, \bibinfo
  {author} {\bibfnamefont {A.}~\bibnamefont {Isaeva}}, \bibinfo {author}
  {\bibfnamefont {T.}~\bibnamefont {Doert}}, \bibinfo {author} {\bibfnamefont
  {A.}~\bibnamefont {Banerjee}}, \bibinfo {author} {\bibfnamefont
  {P.}~\bibnamefont {Lampen-Kelley}}, \bibinfo {author} {\bibfnamefont {D.~G.}\
  \bibnamefont {Mandrus}}, \bibinfo {author} {\bibfnamefont {S.~E.}\
  \bibnamefont {Nagler}}, \bibinfo {author} {\bibfnamefont {J.}~\bibnamefont
  {Sears}}, \bibinfo {author} {\bibfnamefont {Y.-J.}\ \bibnamefont {Kim}},
  \bibinfo {author} {\bibfnamefont {B.}~\bibnamefont {B\"uchner}},\ and\
  \bibinfo {author} {\bibfnamefont {C.}~\bibnamefont {Hess}},\ }\bibfield
  {title} {\bibinfo {title} {\textit{{Unusual Phonon Heat Transport in
  $\ensuremath{\alpha}\text{\ensuremath{-}}{\mathrm{RuCl}}_{3}$: Strong
  Spin-Phonon Scattering and Field-Induced Spin Gap}}},\ }\href
  {https://doi.org/10.1103/PhysRevLett.120.117204} {\bibfield  {journal}
  {\bibinfo  {journal} {Phys. Rev. Lett.}\ }\textbf {\bibinfo {volume} {120}},\
  \bibinfo {pages} {117204} (\bibinfo {year} {2018})}\BibitemShut {NoStop}%
\bibitem [{\citenamefont {Li}\ \emph {et~al.}(2021{\natexlab{b}})\citenamefont
  {Li}, \citenamefont {Zhang}, \citenamefont {Said}, \citenamefont {Fabbris},
  \citenamefont {Mazzone}, \citenamefont {Yan}, \citenamefont {Mandrus},
  \citenamefont {Halász}, \citenamefont {Okamoto}, \citenamefont {Murakami},
  \citenamefont {Dean}, \citenamefont {Lee},\ and\ \citenamefont
  {Miao}}]{li_giant_2021}%
  \BibitemOpen
  \bibfield  {author} {\bibinfo {author} {\bibfnamefont {H.}~\bibnamefont
  {Li}}, \bibinfo {author} {\bibfnamefont {T.~T.}\ \bibnamefont {Zhang}},
  \bibinfo {author} {\bibfnamefont {A.}~\bibnamefont {Said}}, \bibinfo {author}
  {\bibfnamefont {G.}~\bibnamefont {Fabbris}}, \bibinfo {author} {\bibfnamefont
  {D.~G.}\ \bibnamefont {Mazzone}}, \bibinfo {author} {\bibfnamefont {J.~Q.}\
  \bibnamefont {Yan}}, \bibinfo {author} {\bibfnamefont {D.}~\bibnamefont
  {Mandrus}}, \bibinfo {author} {\bibfnamefont {G.~B.}\ \bibnamefont
  {Halász}}, \bibinfo {author} {\bibfnamefont {S.}~\bibnamefont {Okamoto}},
  \bibinfo {author} {\bibfnamefont {S.}~\bibnamefont {Murakami}}, \bibinfo
  {author} {\bibfnamefont {M.~P.~M.}\ \bibnamefont {Dean}}, \bibinfo {author}
  {\bibfnamefont {H.~N.}\ \bibnamefont {Lee}},\ and\ \bibinfo {author}
  {\bibfnamefont {H.}~\bibnamefont {Miao}},\ }\bibfield  {title} {\bibinfo
  {title} {\textit{{Giant Phonon Anomalies in the Proximate {Kitaev} Quantum
  Spin Liquid \ensuremath{\alpha}-{\ensuremath{\mathrm{RuCl}_3}}}}},\ }\href
  {https://doi.org/10.1038/s41467-021-23826-1} {\bibfield  {journal} {\bibinfo
  {journal} {Nat. Commun.}\ }\textbf {\bibinfo {volume} {12}},\ \bibinfo
  {pages} {3513} (\bibinfo {year} {2021}{\natexlab{b}})}\BibitemShut {NoStop}%
\bibitem [{\citenamefont {Reschke}\ \emph {et~al.}(2019)\citenamefont
  {Reschke}, \citenamefont {Tsurkan}, \citenamefont {Do}, \citenamefont {Choi},
  \citenamefont {Lunkenheimer}, \citenamefont {Wang},\ and\ \citenamefont
  {Loidl}}]{reschke_terahertz_2019}%
  \BibitemOpen
  \bibfield  {author} {\bibinfo {author} {\bibfnamefont {S.}~\bibnamefont
  {Reschke}}, \bibinfo {author} {\bibfnamefont {V.}~\bibnamefont {Tsurkan}},
  \bibinfo {author} {\bibfnamefont {S.-H.}\ \bibnamefont {Do}}, \bibinfo
  {author} {\bibfnamefont {K.-Y.}\ \bibnamefont {Choi}}, \bibinfo {author}
  {\bibfnamefont {P.}~\bibnamefont {Lunkenheimer}}, \bibinfo {author}
  {\bibfnamefont {Z.}~\bibnamefont {Wang}},\ and\ \bibinfo {author}
  {\bibfnamefont {A.}~\bibnamefont {Loidl}},\ }\bibfield  {title} {\bibinfo
  {title} {\textit{{Terahertz Excitations in
  $\ensuremath{\alpha}\text{\ensuremath{-}}\mathrm{RuC}{\mathrm{l}}_{3}$:
  Majorana Fermions and Rigid-Plane Shear and Compression Modes}}},\ }\href
  {https://doi.org/10.1103/PhysRevB.100.100403} {\bibfield  {journal} {\bibinfo
   {journal} {Phys. Rev. B}\ }\textbf {\bibinfo {volume} {100}},\ \bibinfo
  {pages} {100403} (\bibinfo {year} {2019})}\BibitemShut {NoStop}%
\bibitem [{\citenamefont {Bruin}\ \emph {et~al.}(2022)\citenamefont {Bruin},
  \citenamefont {Claus}, \citenamefont {Matsumoto}, \citenamefont {Kurita},
  \citenamefont {Tanaka},\ and\ \citenamefont
  {Takagi}}]{bruin_robustness_2022}%
  \BibitemOpen
  \bibfield  {author} {\bibinfo {author} {\bibfnamefont {J.~A.~N.}\
  \bibnamefont {Bruin}}, \bibinfo {author} {\bibfnamefont {R.~R.}\ \bibnamefont
  {Claus}}, \bibinfo {author} {\bibfnamefont {Y.}~\bibnamefont {Matsumoto}},
  \bibinfo {author} {\bibfnamefont {N.}~\bibnamefont {Kurita}}, \bibinfo
  {author} {\bibfnamefont {H.}~\bibnamefont {Tanaka}},\ and\ \bibinfo {author}
  {\bibfnamefont {H.}~\bibnamefont {Takagi}},\ }\bibfield  {title} {\bibinfo
  {title} {\textit{{Robustness of the Thermal {Hall} Effect Close to
  Half-Quantization in \ensuremath{\alpha}-{\ensuremath{\mathrm{RuCl}_3}}}}},\
  }\href {https://doi.org/10.1038/s41567-021-01501-y} {\bibfield  {journal}
  {\bibinfo  {journal} {Nat. Phys.}\ }\textbf {\bibinfo {volume} {18}},\
  \bibinfo {pages} {401} (\bibinfo {year} {2022})}\BibitemShut {NoStop}%
\bibitem [{\citenamefont {Yamashita}\ \emph {et~al.}(2020)\citenamefont
  {Yamashita}, \citenamefont {Gouchi}, \citenamefont {Uwatoko}, \citenamefont
  {Kurita},\ and\ \citenamefont {Tanaka}}]{yamashita_sample_2020}%
  \BibitemOpen
  \bibfield  {author} {\bibinfo {author} {\bibfnamefont {M.}~\bibnamefont
  {Yamashita}}, \bibinfo {author} {\bibfnamefont {J.}~\bibnamefont {Gouchi}},
  \bibinfo {author} {\bibfnamefont {Y.}~\bibnamefont {Uwatoko}}, \bibinfo
  {author} {\bibfnamefont {N.}~\bibnamefont {Kurita}},\ and\ \bibinfo {author}
  {\bibfnamefont {H.}~\bibnamefont {Tanaka}},\ }\bibfield  {title} {\bibinfo
  {title} {\textit{{Sample Dependence of Half-Integer Quantized Thermal Hall
  Effect in the Kitaev Spin-Liquid Candidate
  $\ensuremath{\alpha}\text{\ensuremath{-}}{\mathrm{RuCl}}_{3}$}}},\ }\href
  {https://doi.org/10.1103/PhysRevB.102.220404} {\bibfield  {journal} {\bibinfo
   {journal} {Phys. Rev. B}\ }\textbf {\bibinfo {volume} {102}},\ \bibinfo
  {pages} {220404} (\bibinfo {year} {2020})}\BibitemShut {NoStop}%
\bibitem [{\citenamefont {Lüthi}(2005)}]{luthi_physical_2005}%
  \BibitemOpen
  \bibfield  {author} {\bibinfo {author} {\bibfnamefont {B.}~\bibnamefont
  {Lüthi}},\ }\href@noop {} {\emph {\bibinfo {title} {Physical {Acoustics} in
  the {Solid} {State}}}}\ (\bibinfo {year} {2005})\BibitemShut {NoStop}%
\bibitem [{\citenamefont {Truell}\ \emph {et~al.}(1996)\citenamefont {Truell},
  \citenamefont {Elbaum},\ and\ \citenamefont
  {Chick}}]{truell_ultrasonic_1969}%
  \BibitemOpen
  \bibfield  {author} {\bibinfo {author} {\bibfnamefont {R.}~\bibnamefont
  {Truell}}, \bibinfo {author} {\bibfnamefont {C.}~\bibnamefont {Elbaum}},\
  and\ \bibinfo {author} {\bibfnamefont {B.~B.}\ \bibnamefont {Chick}},\
  }\href@noop {} {\emph {\bibinfo {title} {Ultrasonic {Methods} in {Solid}
  {State} {Physics}}}}\ (\bibinfo {year} {1996})\BibitemShut {NoStop}%
\bibitem [{\citenamefont {Reschke}\ \emph {et~al.}(2018)\citenamefont
  {Reschke}, \citenamefont {Mayr}, \citenamefont {Widmann}, \citenamefont {von
  Nidda}, \citenamefont {Tsurkan}, \citenamefont {Eremin}, \citenamefont {Do},
  \citenamefont {Choi}, \citenamefont {Wang},\ and\ \citenamefont
  {Loidl}}]{reschke_sub-gap_2018}%
  \BibitemOpen
  \bibfield  {author} {\bibinfo {author} {\bibfnamefont {S.}~\bibnamefont
  {Reschke}}, \bibinfo {author} {\bibfnamefont {F.}~\bibnamefont {Mayr}},
  \bibinfo {author} {\bibfnamefont {S.}~\bibnamefont {Widmann}}, \bibinfo
  {author} {\bibfnamefont {H.-A.~K.}\ \bibnamefont {von Nidda}}, \bibinfo
  {author} {\bibfnamefont {V.}~\bibnamefont {Tsurkan}}, \bibinfo {author}
  {\bibfnamefont {M.~V.}\ \bibnamefont {Eremin}}, \bibinfo {author}
  {\bibfnamefont {S.-H.}\ \bibnamefont {Do}}, \bibinfo {author} {\bibfnamefont
  {K.-Y.}\ \bibnamefont {Choi}}, \bibinfo {author} {\bibfnamefont
  {Z.}~\bibnamefont {Wang}},\ and\ \bibinfo {author} {\bibfnamefont
  {A.}~\bibnamefont {Loidl}},\ }\bibfield  {title} {\bibinfo {title}
  {\textit{{Sub-Gap Optical Response in the {Kitaev} Spin-Liquid Candidate
  \ensuremath{\alpha}-\ensuremath{\mathrm{RuCl}_3}}}},\ }\href
  {https://doi.org/10.1088/1361-648X/aae805} {\bibfield  {journal} {\bibinfo
  {journal} {J. Phys.: Condens. Matter}\ }\textbf {\bibinfo {volume} {30}},\
  \bibinfo {pages} {475604} (\bibinfo {year} {2018})}\BibitemShut {NoStop}%
\bibitem [{\citenamefont {Cao}\ \emph {et~al.}(2016)\citenamefont {Cao},
  \citenamefont {Banerjee}, \citenamefont {Yan}, \citenamefont {Bridges},
  \citenamefont {Lumsden}, \citenamefont {Mandrus}, \citenamefont {Tennant},
  \citenamefont {Chakoumakos},\ and\ \citenamefont
  {Nagler}}]{cao_low-temperature_2016}%
  \BibitemOpen
  \bibfield  {author} {\bibinfo {author} {\bibfnamefont {H.~B.}\ \bibnamefont
  {Cao}}, \bibinfo {author} {\bibfnamefont {A.}~\bibnamefont {Banerjee}},
  \bibinfo {author} {\bibfnamefont {J.-Q.}\ \bibnamefont {Yan}}, \bibinfo
  {author} {\bibfnamefont {C.~A.}\ \bibnamefont {Bridges}}, \bibinfo {author}
  {\bibfnamefont {M.~D.}\ \bibnamefont {Lumsden}}, \bibinfo {author}
  {\bibfnamefont {D.~G.}\ \bibnamefont {Mandrus}}, \bibinfo {author}
  {\bibfnamefont {D.~A.}\ \bibnamefont {Tennant}}, \bibinfo {author}
  {\bibfnamefont {B.~C.}\ \bibnamefont {Chakoumakos}},\ and\ \bibinfo {author}
  {\bibfnamefont {S.~E.}\ \bibnamefont {Nagler}},\ }\bibfield  {title}
  {\bibinfo {title} {\textit{{Low-Temperature Crystal and Magnetic Structure of
  $\ensuremath{\alpha}\ensuremath{-}{\mathrm{RuCl}}_{3}$}}},\ }\href
  {https://doi.org/10.1103/PhysRevB.93.134423} {\bibfield  {journal} {\bibinfo
  {journal} {Phys. Rev. B}\ }\textbf {\bibinfo {volume} {93}},\ \bibinfo
  {pages} {134423} (\bibinfo {year} {2016})}\BibitemShut {NoStop}%
\bibitem [{\citenamefont {Lebert}\ \emph {et~al.}(2022)\citenamefont {Lebert},
  \citenamefont {Kim}, \citenamefont {Prishchenko}, \citenamefont {Tsirlin},
  \citenamefont {Said}, \citenamefont {Alatas},\ and\ \citenamefont
  {Kim}}]{Lebert2022}%
  \BibitemOpen
  \bibfield  {author} {\bibinfo {author} {\bibfnamefont {B.~W.}\ \bibnamefont
  {Lebert}}, \bibinfo {author} {\bibfnamefont {S.}~\bibnamefont {Kim}},
  \bibinfo {author} {\bibfnamefont {D.~A.}\ \bibnamefont {Prishchenko}},
  \bibinfo {author} {\bibfnamefont {A.~A.}\ \bibnamefont {Tsirlin}}, \bibinfo
  {author} {\bibfnamefont {A.~H.}\ \bibnamefont {Said}}, \bibinfo {author}
  {\bibfnamefont {A.}~\bibnamefont {Alatas}},\ and\ \bibinfo {author}
  {\bibfnamefont {Y.-J.}\ \bibnamefont {Kim}},\ }\bibfield  {title} {\bibinfo
  {title} {\textit{{Acoustic Phonon Dispersion of
  $\ensuremath{\alpha}\text{\ensuremath{-}}{\mathrm{RuCl}}_{3}$}}},\ }\href
  {https://doi.org/10.1103/PhysRevB.106.L041102} {\bibfield  {journal}
  {\bibinfo  {journal} {Phys. Rev. B}\ }\textbf {\bibinfo {volume} {106}},\
  \bibinfo {pages} {L041102} (\bibinfo {year} {2022})}\BibitemShut {NoStop}%
\bibitem [{\citenamefont {Zherlitsyn}\ \emph {et~al.}(2014)\citenamefont
  {Zherlitsyn}, \citenamefont {Yasin}, \citenamefont {Wosnitza}, \citenamefont
  {Zvyagin}, \citenamefont {Andreev},\ and\ \citenamefont
  {Tsurkan}}]{zherlitsyn_2014}%
  \BibitemOpen
  \bibfield  {author} {\bibinfo {author} {\bibfnamefont {S.}~\bibnamefont
  {Zherlitsyn}}, \bibinfo {author} {\bibfnamefont {S.~S.}\ \bibnamefont
  {Yasin}}, \bibinfo {author} {\bibfnamefont {J.}~\bibnamefont {Wosnitza}},
  \bibinfo {author} {\bibfnamefont {A.}~\bibnamefont {Zvyagin}}, \bibinfo
  {author} {\bibfnamefont {A.}~\bibnamefont {Andreev}},\ and\ \bibinfo {author}
  {\bibfnamefont {V.}~\bibnamefont {Tsurkan}},\ }\bibfield  {title} {\bibinfo
  {title} {\textit{{Spin-Lattice Effects in Selected Antiferromagnetic
  Materials}}},\ }\href@noop {} {\bibfield  {journal} {\bibinfo  {journal} {Low
  Temp. Phys.}\ }\textbf {\bibinfo {volume} {40}},\ \bibinfo {pages} {123}
  (\bibinfo {year} {2014})}\BibitemShut {NoStop}%
\bibitem [{Lut()}]{Luthi1970}%
  \BibitemOpen
  \href@noop {} {\bibinfo  {journal} {B. L\"uthi, T. J. Moran, R. J. Pollina,
  \textit{Elastic and Magnetoelastic Effects in Rare Earth Metals}, J. Phys.
  Chem. Solids {\bf 31}, 1735 (1970), ibid: \textit{Sound Propagation Near
  Magnetic Phase Transitions}, 1741 (1970)}\ }\BibitemShut {NoStop}%
\bibitem [{Gla()}]{Glamazda2017}%
  \BibitemOpen
\bibfield  {journal} {  }\href@noop {} {\bibinfo  {journal} {A. Glamazda, P.
  Lemmens, S.-H. Do, Y. S. Kwon, and K.-Y. Choi, \textit{Relation between
  Kitaev Magnetism and Structure in
  $\ensuremath{\alpha}\text{\ensuremath{-}}{\mathrm{RuCl}}_{3}$}, Phys. Rev. B
  {\bf 95}, 174429 (2017)}\ }\BibitemShut {NoStop}%
\bibitem [{Par()}]{Park2016}%
  \BibitemOpen
\bibfield  {journal} {  }\href@noop {} {\bibinfo  {journal} {S.-Y. Park, S.-H.
  Do, K.-Y. Choi, D. Jang, T.-H. Jang, J. Schefer, C.-M. Wu, J. S. Gardner, J.
  M. S. Park, J.-H. Park, and S. Ji, \textit{Emergence of the Isotropic Kitaev
  Honeycomb Lattice with Two-Dimensional Ising Universality in
  $\ensuremath{\alpha}\text{\ensuremath{-}}{\mathrm{RuCl}}_{3}$},
  arXiv:1609.05690}\ }\BibitemShut {NoStop}%
\bibitem [{li_()}]{li_divergence_2021}%
  \BibitemOpen
\bibfield  {journal} {  }\href@noop {} {\bibinfo  {journal} {H. Li, A. Said, J.
  Q. Yan, D. M. Mandrus, H. N. Lee, S. Okamoto, G. B. Halász, and H. Miao,
  \textit{Divergence of Majorana-Phonon Scattering in Kitaev Quantum Spin
  Liquid}, arXiv:2112.02015}\ }\BibitemShut {NoStop}%
\bibitem [{Nas()}]{Nasu2015}%
  \BibitemOpen
\bibfield  {journal} {  }\href@noop {} {\bibinfo  {journal} {J. Nasu, M.
  Udagawa, and Y. Motome, \textit{Thermal Fractionalization of Quantum Spins in
  a Kitaev Model: Temperature-Linear Specific Heat and Coherent Transport of
  Majorana Fermions}, Phys. Rev. B {\bf 92}, 115122 (2015)}\ }\BibitemShut
  {NoStop}%
\bibitem [{Met()}]{Metavitsiadis2017}%
  \BibitemOpen
\bibfield  {journal} {  }\href@noop {} {\bibinfo  {journal} {A. Metavitsiadis,
  A. Pidatella, and W. Brenig, \textit{Thermal Transport in a Two-Dimensional
  $\mathbb{Z}_2$ Spin Liquid}, Phys. Rev. B {\bf 96}, 205121 (2017)}\
  }\BibitemShut {NoStop}%
\bibitem [{\citenamefont {Singh}\ \emph {et~al.}()\citenamefont {Singh},
  \citenamefont {Stavropoulos},\ and\ \citenamefont {Perkins}}]{Susmita2023}%
  \BibitemOpen
\bibfield  {journal} {  }\bibfield  {author} {\bibinfo {author} {\bibfnamefont
  {S.}~\bibnamefont {Singh}}, \bibinfo {author} {\bibfnamefont {P.~P.}\
  \bibnamefont {Stavropoulos}},\ and\ \bibinfo {author} {\bibfnamefont {N.~B.}\
  \bibnamefont {Perkins}},\ }\bibfield  {title} {\bibinfo {title}
  {\textit{{Phonon} {Dynamics} in the {Generalized} {Kitaev} {Spin}
  {Liquid}}},\ }\href@noop {} {\bibinfo  {journal} {arxiv:2302.01254}\
  }\BibitemShut {NoStop}%
\bibitem [{\citenamefont {Wang}\ \emph {et~al.}(2017)\citenamefont {Wang},
  \citenamefont {Reschke}, \citenamefont {Hüvonen}, \citenamefont {Do},
  \citenamefont {Choi}, \citenamefont {Gensch}, \citenamefont {Nagel},
  \citenamefont {Rõõm},\ and\ \citenamefont {Loidl}}]{wang_magnetic_2017}%
  \BibitemOpen
\bibfield  {journal} {  }\bibfield  {author} {\bibinfo {author} {\bibfnamefont
  {Z.}~\bibnamefont {Wang}}, \bibinfo {author} {\bibfnamefont {S.}~\bibnamefont
  {Reschke}}, \bibinfo {author} {\bibfnamefont {D.}~\bibnamefont {Hüvonen}},
  \bibinfo {author} {\bibfnamefont {S.-H.}\ \bibnamefont {Do}}, \bibinfo
  {author} {\bibfnamefont {K.-Y.}\ \bibnamefont {Choi}}, \bibinfo {author}
  {\bibfnamefont {M.}~\bibnamefont {Gensch}}, \bibinfo {author} {\bibfnamefont
  {U.}~\bibnamefont {Nagel}}, \bibinfo {author} {\bibfnamefont
  {T.}~\bibnamefont {Rõõm}},\ and\ \bibinfo {author} {\bibfnamefont
  {A.}~\bibnamefont {Loidl}},\ }\bibfield  {title} {\bibinfo {title}
  {\textit{{Magnetic {Excitations} and {Continuum} of a {Possibly}
  {Field}-{Induced} {Quantum} {Spin} {Liquid} in
  $\ensuremath{\alpha}\text{\ensuremath{-}}{\mathrm{RuCl}}_{3}$}}},\ }\href
  {https://doi.org/10.1103/PhysRevLett.119.227202} {\bibfield  {journal}
  {\bibinfo  {journal} {Phys. Rev. Lett.}\ }\textbf {\bibinfo {volume} {119}},\
  \bibinfo {pages} {227202} (\bibinfo {year} {2017})}\BibitemShut {NoStop}%
\bibitem [{tsh()}]{tsh}%
  \BibitemOpen
  \href@noop {} {\bibinfo  {journal} {Using the {\it perturbative} finite-field
  Hamiltonian \cite{kitaev_anyons_2006}, $T^\star(H)$ will increases slightly
  with $H$. At $H_c$, however, for $\kappa\sim 0.04$, this effect is negligible
  \cite{Motome2019}. Treating $H$ non-perturbatively, $T^\star(H)$ will
  decrease, because of flux mobility \cite{Trebst2022}. I.\,e.,
  $T^\star(H_c)\lesssim \unit[3]{K}$ is certainly satisfied}\ }\BibitemShut
  {NoStop}%
\bibitem [{\citenamefont {Knolle}\ \emph {et~al.}(2019)\citenamefont {Knolle},
  \citenamefont {Moessner},\ and\ \citenamefont {Perkins}}]{Knolle2019}%
  \BibitemOpen
\bibfield  {journal} {  }\bibfield  {author} {\bibinfo {author} {\bibfnamefont
  {J.}~\bibnamefont {Knolle}}, \bibinfo {author} {\bibfnamefont
  {R.}~\bibnamefont {Moessner}},\ and\ \bibinfo {author} {\bibfnamefont
  {N.~B.}\ \bibnamefont {Perkins}},\ }\bibfield  {title} {\bibinfo {title}
  {\textit{{Bond}-{Disordered} {Spin} {Liquid} and the {Honeycomb} {Iridate}
  ${\mathrm{H}}_{3}{\mathrm{LiIr}}_{2}{\mathrm{O}}_{6}$: {Abundant}
  {Low}-{Energy} {Density} of {States} from {Random} {Majorana} {Hopping}}},\
  }\href {https://doi.org/10.1103/PhysRevLett.122.047202} {\bibfield  {journal}
  {\bibinfo  {journal} {Phys. Rev. Lett.}\ }\textbf {\bibinfo {volume} {122}},\
  \bibinfo {pages} {047202} (\bibinfo {year} {2019})}\BibitemShut {NoStop}%
\bibitem [{\citenamefont {Kao}\ \emph {et~al.}(2021)\citenamefont {Kao},
  \citenamefont {Knolle}, \citenamefont {Hal\'asz}, \citenamefont {Moessner},\
  and\ \citenamefont {Perkins}}]{Kao2021}%
  \BibitemOpen
  \bibfield  {author} {\bibinfo {author} {\bibfnamefont {W.-H.}\ \bibnamefont
  {Kao}}, \bibinfo {author} {\bibfnamefont {J.}~\bibnamefont {Knolle}},
  \bibinfo {author} {\bibfnamefont {G.~B.}\ \bibnamefont {Hal\'asz}}, \bibinfo
  {author} {\bibfnamefont {R.}~\bibnamefont {Moessner}},\ and\ \bibinfo
  {author} {\bibfnamefont {N.~B.}\ \bibnamefont {Perkins}},\ }\bibfield
  {title} {\bibinfo {title} {\textit{{Vacancy}-{Induced} {Low}-{Energy}
  {Density} of {States} in the {Kitaev} {Spin} {Liquid}}},\ }\href
  {https://doi.org/10.1103/PhysRevX.11.011034} {\bibfield  {journal} {\bibinfo
  {journal} {Phys. Rev. X}\ }\textbf {\bibinfo {volume} {11}},\ \bibinfo
  {pages} {011034} (\bibinfo {year} {2021})}\BibitemShut {NoStop}%
\bibitem [{\citenamefont {Kao}\ and\ \citenamefont
  {Perkins}(2021)}]{Kao2021localization}%
  \BibitemOpen
  \bibfield  {author} {\bibinfo {author} {\bibfnamefont {W.-H.}\ \bibnamefont
  {Kao}}\ and\ \bibinfo {author} {\bibfnamefont {N.~B.}\ \bibnamefont
  {Perkins}},\ }\bibfield  {title} {\bibinfo {title} {\textit{{Disorder} upon
  {Disorder}: {Localization} {Effects} in the {Kitaev} {Spin} {Liquid}}},\
  }\href {https://doi.org/https://doi.org/10.1016/j.aop.2021.168506} {\bibfield
   {journal} {\bibinfo  {journal} {Ann. Phys.}\ }\textbf {\bibinfo {volume}
  {435}},\ \bibinfo {pages} {168506} (\bibinfo {year} {2021})}\BibitemShut
  {NoStop}%
\bibitem [{\citenamefont {Dantas}\ and\ \citenamefont
  {Andrade}(2022)}]{Vitor2022}%
  \BibitemOpen
  \bibfield  {author} {\bibinfo {author} {\bibfnamefont {V.}~\bibnamefont
  {Dantas}}\ and\ \bibinfo {author} {\bibfnamefont {E.~C.}\ \bibnamefont
  {Andrade}},\ }\bibfield  {title} {\bibinfo {title} {\textit{{Disorder},
  {Low}-{Energy} {Excitations}, and {Topology} in the {Kitaev} {Spin}
  {Liquid}}},\ }\href {https://doi.org/10.1103/PhysRevLett.129.037204}
  {\bibfield  {journal} {\bibinfo  {journal} {Phys. Rev. Lett.}\ }\textbf
  {\bibinfo {volume} {129}},\ \bibinfo {pages} {037204} (\bibinfo {year}
  {2022})}\BibitemShut {NoStop}%
\bibitem [{\citenamefont {Balz}\ \emph {et~al.}(2019)\citenamefont {Balz},
  \citenamefont {Lampen-Kelley}, \citenamefont {Banerjee}, \citenamefont {Yan},
  \citenamefont {Lu}, \citenamefont {Hu}, \citenamefont {Yadav}, \citenamefont
  {Takano}, \citenamefont {Liu}, \citenamefont {Tennant}, \citenamefont
  {Lumsden}, \citenamefont {Mandrus},\ and\ \citenamefont
  {Nagler}}]{balz_phasediagram_2019}%
  \BibitemOpen
  \bibfield  {author} {\bibinfo {author} {\bibfnamefont {C.}~\bibnamefont
  {Balz}}, \bibinfo {author} {\bibfnamefont {P.}~\bibnamefont {Lampen-Kelley}},
  \bibinfo {author} {\bibfnamefont {A.}~\bibnamefont {Banerjee}}, \bibinfo
  {author} {\bibfnamefont {J.}~\bibnamefont {Yan}}, \bibinfo {author}
  {\bibfnamefont {Z.}~\bibnamefont {Lu}}, \bibinfo {author} {\bibfnamefont
  {X.}~\bibnamefont {Hu}}, \bibinfo {author} {\bibfnamefont {S.~M.}\
  \bibnamefont {Yadav}}, \bibinfo {author} {\bibfnamefont {Y.}~\bibnamefont
  {Takano}}, \bibinfo {author} {\bibfnamefont {Y.}~\bibnamefont {Liu}},
  \bibinfo {author} {\bibfnamefont {D.~A.}\ \bibnamefont {Tennant}}, \bibinfo
  {author} {\bibfnamefont {M.~D.}\ \bibnamefont {Lumsden}}, \bibinfo {author}
  {\bibfnamefont {D.}~\bibnamefont {Mandrus}},\ and\ \bibinfo {author}
  {\bibfnamefont {S.~E.}\ \bibnamefont {Nagler}},\ }\bibfield  {title}
  {\bibinfo {title} {\textit{{Finite Field Regime for a Quantum Spin Liquid in
  $\ensuremath{\alpha}\text{\ensuremath{-}}{\mathrm{RuCl}}_{3}$}}},\ }\href
  {https://doi.org/10.1103/PhysRevB.100.060405} {\bibfield  {journal} {\bibinfo
   {journal} {Phys. Rev. B}\ }\textbf {\bibinfo {volume} {100}},\ \bibinfo
  {pages} {060405} (\bibinfo {year} {2019})}\BibitemShut {NoStop}%
\bibitem [{\citenamefont {Sears}\ \emph {et~al.}(2017)\citenamefont {Sears},
  \citenamefont {Zhao}, \citenamefont {Xu}, \citenamefont {Lynn},\ and\
  \citenamefont {Kim}}]{Sears2017}%
  \BibitemOpen
  \bibfield  {author} {\bibinfo {author} {\bibfnamefont {J.~A.}\ \bibnamefont
  {Sears}}, \bibinfo {author} {\bibfnamefont {Y.}~\bibnamefont {Zhao}},
  \bibinfo {author} {\bibfnamefont {Z.}~\bibnamefont {Xu}}, \bibinfo {author}
  {\bibfnamefont {J.~W.}\ \bibnamefont {Lynn}},\ and\ \bibinfo {author}
  {\bibfnamefont {Y.-J.}\ \bibnamefont {Kim}},\ }\bibfield  {title} {\bibinfo
  {title} {\textit{{Phase Diagram of
  $\ensuremath{\alpha}\ensuremath{-}{\mathrm{RuCl}}_{3}$ in an In-Plane
  Magnetic Field}}},\ }\href {https://doi.org/10.1103/PhysRevB.95.180411}
  {\bibfield  {journal} {\bibinfo  {journal} {Phys. Rev. B}\ }\textbf {\bibinfo
  {volume} {95}},\ \bibinfo {pages} {180411} (\bibinfo {year}
  {2017})}\BibitemShut {NoStop}%
\bibitem [{\citenamefont {Wolter}\ \emph {et~al.}(2017)\citenamefont {Wolter},
  \citenamefont {Corredor}, \citenamefont {Janssen}, \citenamefont {Nenkov},
  \citenamefont {Sch\"onecker}, \citenamefont {Do}, \citenamefont {Choi},
  \citenamefont {Albrecht}, \citenamefont {Hunger}, \citenamefont {Doert},
  \citenamefont {Vojta},\ and\ \citenamefont {B\"uchner}}]{PhysRevB.96.041405}%
  \BibitemOpen
  \bibfield  {author} {\bibinfo {author} {\bibfnamefont {A.~U.~B.}\
  \bibnamefont {Wolter}}, \bibinfo {author} {\bibfnamefont {L.~T.}\
  \bibnamefont {Corredor}}, \bibinfo {author} {\bibfnamefont {L.}~\bibnamefont
  {Janssen}}, \bibinfo {author} {\bibfnamefont {K.}~\bibnamefont {Nenkov}},
  \bibinfo {author} {\bibfnamefont {S.}~\bibnamefont {Sch\"onecker}}, \bibinfo
  {author} {\bibfnamefont {S.-H.}\ \bibnamefont {Do}}, \bibinfo {author}
  {\bibfnamefont {K.-Y.}\ \bibnamefont {Choi}}, \bibinfo {author}
  {\bibfnamefont {R.}~\bibnamefont {Albrecht}}, \bibinfo {author}
  {\bibfnamefont {J.}~\bibnamefont {Hunger}}, \bibinfo {author} {\bibfnamefont
  {T.}~\bibnamefont {Doert}}, \bibinfo {author} {\bibfnamefont
  {M.}~\bibnamefont {Vojta}},\ and\ \bibinfo {author} {\bibfnamefont
  {B.}~\bibnamefont {B\"uchner}},\ }\bibfield  {title} {\bibinfo {title}
  {\textit{{Field-Induced Quantum Criticality in the Kitaev System
  $\ensuremath{\alpha}\ensuremath{-}{\mathrm{RuCl}}_{3}$}}},\ }\href
  {https://doi.org/10.1103/PhysRevB.96.041405} {\bibfield  {journal} {\bibinfo
  {journal} {Phys. Rev. B}\ }\textbf {\bibinfo {volume} {96}},\ \bibinfo
  {pages} {041405} (\bibinfo {year} {2017})}\BibitemShut {NoStop}%
\end{thebibliography}%

\end{document}